\documentclass[ 
    aps,
    prl,
    twocolumn,
    letterpaper, 
    10pt,
    superscriptaddress, 
    showpacs,
    showkeys,
    notitlepage,
    amsmath, 
    amssymb, 
    floatfix 
]{revtex4-1}

\usepackage{graphicx}
\usepackage{dcolumn}
\usepackage{bm}
\usepackage{color}
\usepackage{amssymb}
\usepackage{amsmath}
\usepackage{hyperref}
\usepackage{bbold}
\usepackage{accents}

\newcommand{\harpvecsign}{\scriptscriptstyle\text{\tiny$\leftrightarrow$}}
\newcommand{\harpoonvec}[2]{%
  \ifx\displaystyle#1\doalign{$\harpvecsign$}{#1#2}\fi
  \ifx\textstyle#1\doalign{$\harpvecsign$}{#1#2}\fi
  \ifx\scriptstyle#1\doalign{\scalebox{.6}[.9]{$\harpvecsign$}}{#1#2}\fi
  \ifx\scriptscriptstyle#1\doalign{\scalebox{.5}[.8]{$\harpvecsign$}}{#1#2}\fi
}
\newcommand{\doalign}[2]{%
 {\vbox{\offinterlineskip\ialign{\hfil##\hfil\cr#1\cr$#2$\cr}}}%
}

\newcommand{\bel}{\begin{equation}}
\newcommand{\eel}{\end{equation}}

\newcommand{\skyp}[1]{}

\newcommand{\fr}{\frac}

\newcommand{\vare}{\varepsilon}

\newcommand{\ee}{\end{equation}}
\newcommand{\be}{\begin{equation}}
\newcommand{\mbf}{\mathbf}

\newcommand{\bal}{\begin{eqnarray} }
\newcommand{\eal}{\end{eqnarray}}
\newcommand{\ba}{\begin{eqnarray*}}
\newcommand{\ea}{\end{eqnarray*}}

\newcommand{\reffig}[1]{Fig.~\ref{#1}}

\newcommand{\ket}[1]{| #1 \rangle}
\newcommand{\bra}[1]{\langle #1 |}

\newcommand{\bp}{{\mathbf p}}
\newcommand{\br}{{\mathbf r}}

\newcommand{\bR}{{\mathbf R}}

\newcommand{\bk}{{\mathbf k}}

\newcommand{\bG}{{\mathbf G}}

\newcommand{\refeq}[1]{Eq.~\eqref{#1}}

\begin{document} 

\title{ Topological quantum optics using atom-like emitter arrays \\
 coupled to photonic crystals}

\author{J. Perczel}
\email[email: ]{jperczel@mit.edu}
\affiliation{Physics Department, Massachusetts Institute of Technology, Cambridge, MA 02139, USA}
\affiliation{Physics Department, Harvard University, Cambridge,
MA 02138, USA}

\author{J. Borregaard}
\affiliation{Physics Department, Harvard University, Cambridge,
MA 02138, USA}
\affiliation{QMATH, Department of Mathematical Sciences, University of Copenhagen, Copenhagen, 2100, Denmark}

\author{D. E. Chang}
\affiliation{ICFO - Institut de Ciencies Fotoniques, The Barcelona Institute of Science and Technology, 08860 Castelldefels, Barcelona, Spain}
\affiliation{ICREA - Instituci\'{o} Catalana de Recerca i Estudis Avan\c{c}ats, 08015 Barcelona, Spain}

\author{S. F. Yelin}
\affiliation{Physics Department, Harvard University, Cambridge,
MA 02138, USA}
\affiliation{Department of Physics, University of Connecticut, Storrs, Connecticut 06269, USA}

\author{M. D. Lukin}
\affiliation{Physics Department, Harvard University, Cambridge,
MA 02138, USA}


\date{\today}

\bigskip
\bigskip
\bigskip
\begin{abstract} 

We propose a nanophotonic platform for topological quantum optics. Our system is composed of a two-dimensional lattice of non-linear quantum emitters with optical transitions embedded in a photonic crystal slab. The emitters interact through the guided modes of the photonic crystal, and a uniform magnetic field gives rise to large topological band gaps and an almost completely flat topological band. Topological edge states arise on the boundaries of the system that are protected by the large gap against missing lattice sites and to the inhomogeneous broadening of emitters. These results pave the way for exploring topological many-body states in quantum optical systems. 


\end{abstract}

\maketitle
 


Since their discovery in electronic systems \cite{Klitzing1980,Tsui1982,Konig2007}, topological phenomena have been explored in a variety of systems, including microwave \cite{Haldane2008,Raghu2008,Wang2008,Liu2012,Wang2009,Yu2008}, photonic \cite{Rechtsman2013b,Hafezi2011,Hafezi2013,Fang2012}, acoustic \cite{Khanikaev2015,Yang2015,Lu2017}, mechanical \cite{Susstrunk2015,Mousavi2015,Nash2015}, and cold atom environments \cite{Dalibard2011,Eckardt2017}. While such systems can exhibit reflection-free excitation transport on system edges, even in the presence of imperfections,  they are generally  robust only with respect to certain types of disorder \cite{Lu2014,Yang2015,Huber2016}. This is in contrast to electronic systems with time-reversal symmetry breaking that are robust to arbitrary perturbations \cite{Laughlin1981,Halperin1982}. In addition, the linear acoustic, microwave or photonic systems generally lack the interactions between constituent particles  required to obtain exotic states of matter such as the fractional quantum Hall effect \cite{Laughlin1983}.

Recently, quantum emitter arrays in free-space have been shown to support robust topological states at optical frequencies \cite{Perczel2017a,Perczel2017b,Bettles2017}. Such systems, however, require deeply subwavelenth interatomic spacing, which is experimentally very challenging to achieve. At the same time, there has been significant interest in combining photonic systems with non-linear quantum emitters to study strongly-correlated states of light and matter \cite{Carusotto2013,Gonzalez-Tudela2015,Douglas2015,Lodahl2017,Angelakis2017,Cirac2017}. Pioneering work at the intersection of these two approaches has demonstrated the coupling of a {\it single} quantum emitter to a topological photonic interface \cite{Barik2018}, but the development of a truly robust, large-scale many-body platform for topological quantum optics remains an outstanding challenge. 
Furthermore, strongly correlated systems must be based on topological bands with negligible dispersion relative to the energy scale of the interactions, which typically requires careful fine-tuning of the system parameters that is difficult to realize in practice \cite{Neupert2011,Tang2011,Sun2011,Peter2015,Yao2012,Yao2013}. Realization of robust topological systems in the optical domain is especially interesting in light of potential applications to quantum networking \cite{Sangouard2011}.

\begin{figure}[h!]
\centering
\includegraphics[width=0.45 \textwidth]{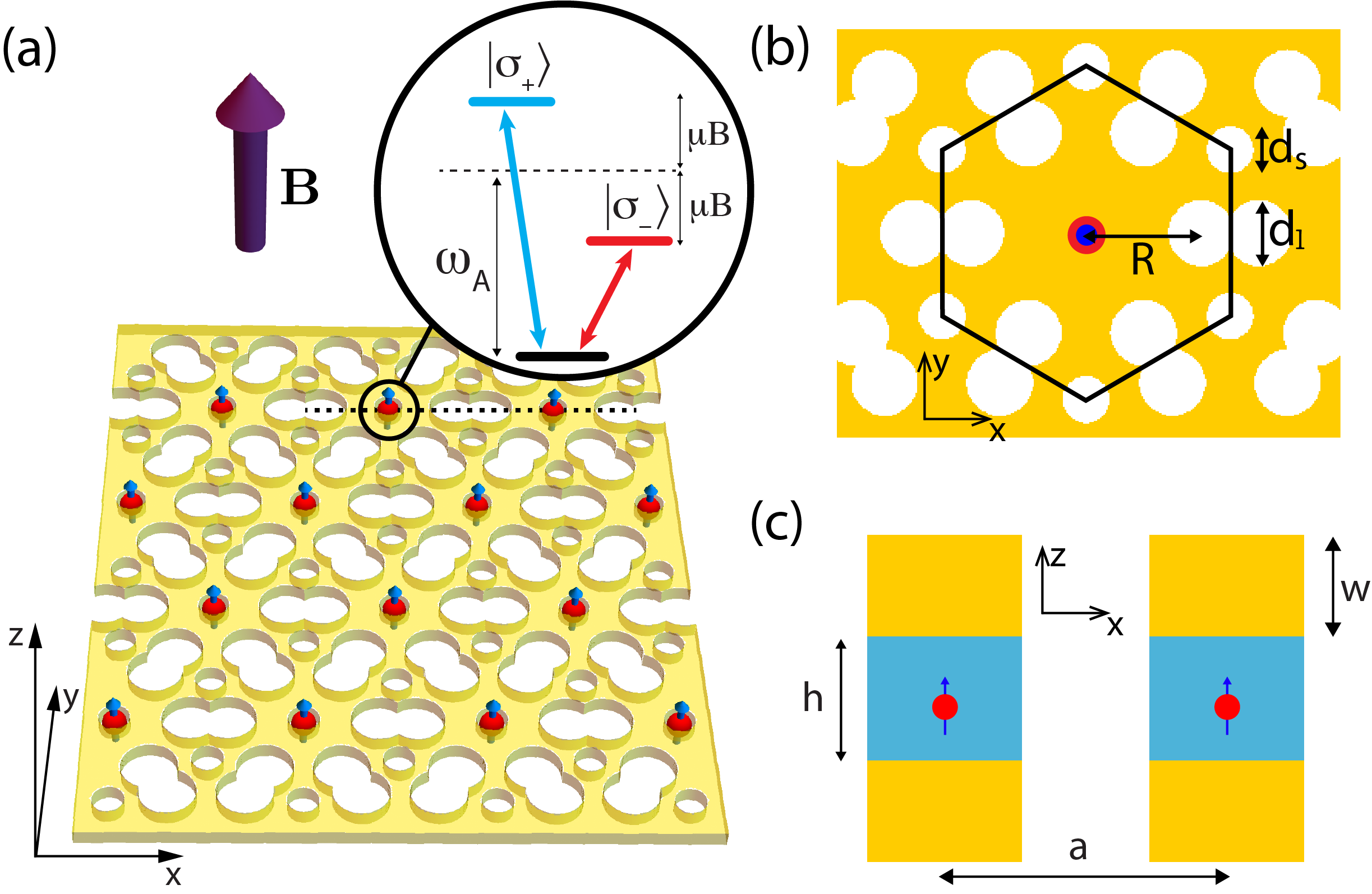}
\caption{
\label{fig:PC_structure}
(a) Schematic depiction of the photonic crystal slab with air holes and the embedded triangular emitter array. Emitter spacing is $a=240$nm, transition wavelength is $\lambda=738$nm. Out-of-plane B-field splits the $\ket{\sigma_+}$ and $\ket{\sigma_-}$ atomic transitions. (b) Unit cell of the photonic crystal (black lines). The emitter is embedded at the center. The diameters of small and large holes are $d_\text{s}=40$nm and $d_\text{l}=56$nm, respectively. The large holes are centered $R=96$nm away from the emitter. (c) Cross section of the photonic structure along the dotted line indicated in (a). The SiVs are embedded in a diamond ($n_\text{d}=2.4$) layer of thickness $h=120$nm, which is surrounded on both sides by GaP ($n_\text{GaP}=3.25$) layers of thickness $w=75$nm.} 
\end{figure} 

In this manuscript, we propose a novel, realistic interface between a two-dimensional photonic crystal and an atomic emitter array, whose combined properties address all of the above issues to realize a robust topological system with strong interactions at optical frequencies. Specifically, we consider a two-dimensional atomic emitter array integrated with a two dimensional photonic crystal slab. We show that in the presence of an out-of-plane magnetic field, the hybridization of the atomic levels and the photonic modes of the slab lead to energy bands with non-trivial Chern numbers. In contrast to free-space realizations, we find very large topological band gaps and the existence of an almost completely flat topological band. This makes the system a strong candidate for the realization of fractional phases, such as fractional Chern insulators \cite{Neupert2011,Tang2011,Sun2011,Peter2015,Yao2012,Yao2013}. We also demonstrate that topological edge states form on the system boundary that are robust to imperfections in the lattice and inhomogeneous broadening of the emitters. Notably, the required lattice spacing of the emitters is comparable to the optical wavelength, which is an order of magnitude larger than for free-space systems \cite{Perczel2017a,Perczel2017b,Bettles2017}. The hybrid approach considered here thus opens up a rich and experimentally accessible platform for exploring topological quantum optics and many-body physics.    

\begin{figure}[h!]
\centering
\includegraphics[width=.45 \textwidth]{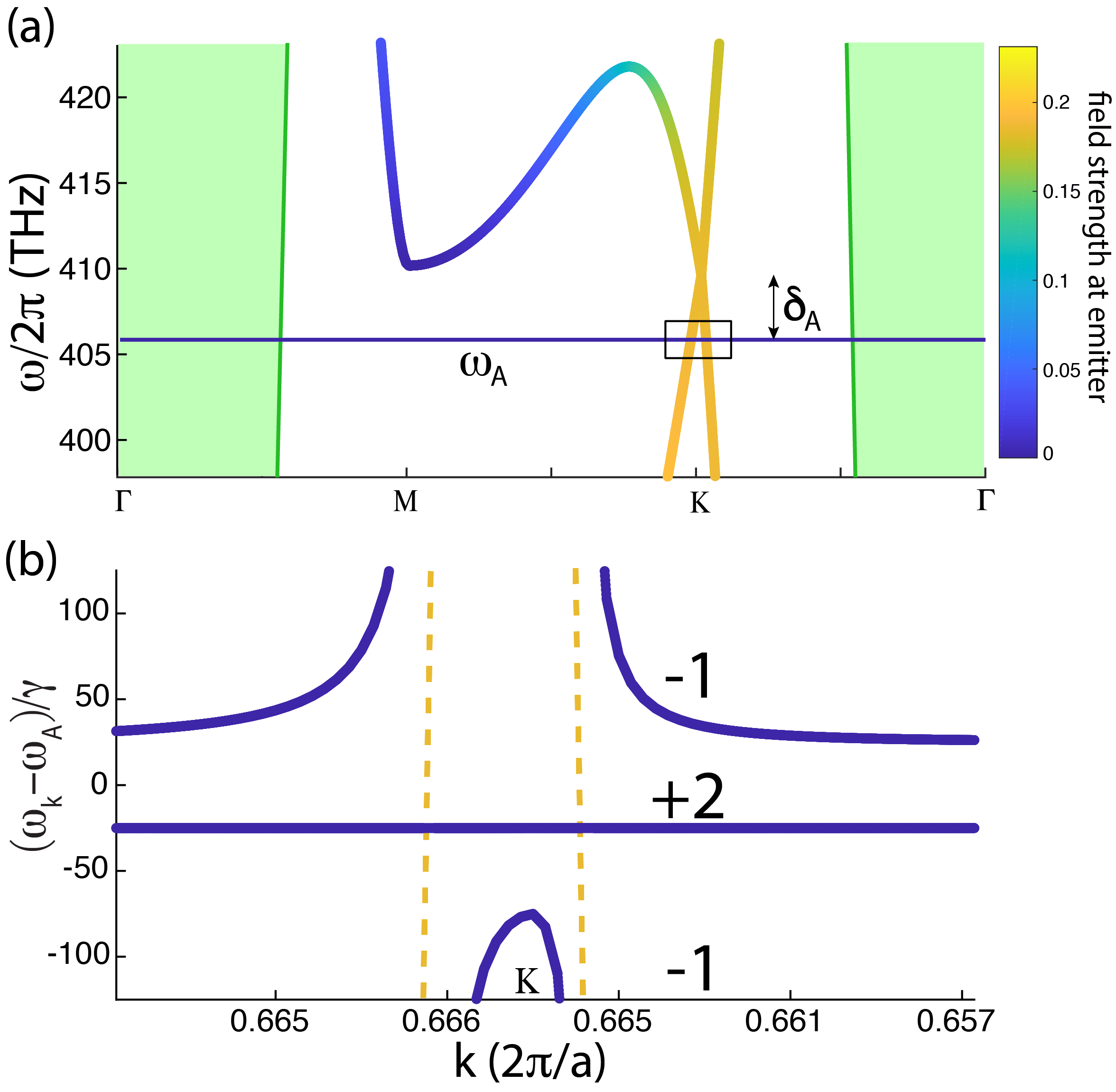}
\caption{
\label{fig:atom_bands}
 (a) TE-like photonic band structure of the slab in the vicinity of the SiV transition frequency $\omega_A$. The light cone region is shaded green. The tip of the Dirac cone is detuned by $\delta_A=3.74$THz from $\omega_A$. Band colors reflect the normalized field strength $\left|\mbf E_\bk(\mbf r_A)\right|^2\!a^3$ at the location of the emitters. (b) Hybrid atomic-photonic bands in the presence of an out-of-plane B-field ($\mu B=25\gamma$) in the immediate vicinity of the $\mbf K$ point (see black box in (a)) for a detuning of $\delta_A=0.321$THz. Yellow dashed lines indicate the bare photonic bands. One atomic band does not interact with the slab modes and remains flat. The other atomic band interacts with the photonic modes and avoided crossings arise. Chern numbers of the bands are indicated by black numbers.} 
\end{figure}

\reffig{fig:PC_structure}(a) provides a schematic depiction of the physical system. A triangular lattice of quantum emitters with spacing $a = \lambda/3$ is embedded in a two-dimensional photonic crystal (PhC) slab of air holes. Each atomic emitter has two optical transitions with wavelength $\lambda$ between the ground state $\ket{g}$ and the two excited states $\ket{\sigma_+}$ and $\ket{\sigma_-}$ and is surrounded by a cavity-like hole structure. The emitters interact primarily via the guided modes of the slab. A uniform out-of-plane magnetic field Zeeman-splits the excited states of each emitter. The resulting hybridized polaritonic bands feature non-trivial Chern numbers and one of the topological bands is almost completely flat (\reffig{fig:atom_bands}(b)). In addition, a large energy gap is formed between the bands, which is two orders of magnitude larger than the gap in free space for comparable emitter spacings \cite{Perczel2017a,Perczel2017b}. Inside the gap topological edge states appear, which are protected by the large gap from scattering into extended bulk states or the guided modes of the slab and have a large group velocity. Thus these modes carry energy around the system boundary rapidly with minimal losses even in the presence of lattice imperfections and inhomogeneous emitter broadening. 

For concreteness, we now focus on the hybrid PhC structure  illustrated in \reffig{fig:PC_structure}. The hexagonal unit cell of the PhC slab made of gallium phosphate (GaP) with air holes is shown in \reffig{fig:PC_structure}(b). Each atomic emitter is placed at the geometric center of the unit cell, forming a periodic triangular lattice. In addition, the emitters are placed in the middle of the slab (in the $z=0$ mirror plane) to ensure that they only couple to TE-like guided modes \cite{Joannopoulos2008}. Such an arrangement can be achieved for different kinds of quantum emitters by introducing slight modifications to the PhC structure. One option is to embed quantum dots directly into the dielectric slab \cite{Javadi2015,Sun2016,Lodahl2015,Arcari2014,Sollner2015,Barik2018}. Another option is to create an additional hole at the center of the cell and trap a neutral atom in the $z=0$ plane using a combination of far-off resonance optical trapping, vacuum forces and side illumination \cite{Gonzalez-Tudela2015} (see also Refs.~\cite{Thompson2013,Tiecke2014,Hood2016,Goban2015,Goban2014,Yu2014}). Alternatively, atom-like color defects in diamond such as Silicon Vacancy  (SiV) color centers can be integrated with the photonic structure \cite{Faraon2012,Riedrich-Moller2014,Sipahigil2016}, by using  
 a thin diamond layer  sandwiched between two layers of GaP as shown in \reffig{fig:PC_structure}(c). 

The TE-like guided bands of the  slab near the emitter frequency are shown in \reffig{fig:atom_bands}(a) \cite{SM_ref}. The colors of the bands indicate the normalized field strength $\left|\mbf E_\bk(\mbf r_A)\right|^2\!a^3$ at the location of the emitter inside the unit cell, where $\mbf E_\bk(\br)$ denotes the classical field solutions of Maxwell's equations for Bloch quasi-momentum $\bk$ \cite{Glauber1991,Perczel2018,Joannopoulos2008,SM_ref}. This photonic structure was specifically designed to ensure that there are no other guided modes within a few THz energy range just below the tip of the Dirac cone. Such a photonic spectrum is a general feature of PhC slabs with the cavity-like hole structure shown in \reffig{fig:PC_structure}(b) and can be achieved for a wide range of geometric and material parameters \cite{dirac_theory_draft}. The thickness of the layers and the size and spacing of the holes of the diamond-GaP structure are chosen such that the tip of the Dirac cone $\omega_\text{Dirac}$ is tuned within a few hundred GHz of $\omega_A=2\pi c/\lambda$, the transition frequency of the SiV emitters ($\lambda = 738$nm). Thus the emitters interact primarily through the guided modes of the Dirac cone.

The dynamics of the embedded emitters (no-jump evolution in the master equation) in the presence of a magnetic field can be described by the following non-Hermitian Hamiltonian \cite{Antezza2009,Shahmoon2017,Bienaime2012,Guerin2016,Perczel2017a,Perczel2017b}
\bal\label{Hamiltonian}
H=\hbar\sum\limits_{i=1}^{N}\sum\limits_{\alpha=\sigma_+,\sigma_-}\left(\omega_A+\text{sgn}(\alpha_i)\mu B-\text{i}\fr{\Gamma}{2}\right)\ket{\alpha_i}\bra{\alpha_i}\nonumber \\
+\fr{3\pi\gamma \hbar c}{\omega_An_\text{d}}\sum\limits_{i\neq j}\sum\limits_{\alpha,\beta=\sigma_+,\sigma_-}G_{\alpha\beta}(\br_i-\br_j)\ket{\alpha_i}\bra{\beta_j},\qquad
\eal
where $N$ is the number of atoms, we define ${\text{sgn}(\sigma_\pm)=\pm}$, $\mu B$ is the Zeeman-shift of the emitters with magnetic moment $\mu$ due to the magnetic field $\mbf{B}=B\hat z$, $\gamma$ is the emission rate of SiVs in bulk diamond, $c$ is the speed of light and $n_\text{d}$ is the refractive index of diamond. The dipolar spin-spin interaction of emitters is described by the dyadic Green's function $G_{\alpha\beta}(\br)$ and $\br_i$ denotes the position of the $i^\text{th}$ atom. The irreversible decay rate of individual emitters inside the slab is given by ${\Gamma=\Gamma_\text{PC}+\Gamma_0}$, where $\Gamma_\text{PC}$ and $\Gamma_0$ account for the coupling of the emitters to the guided slab modes and free-space modes, respectively. In general, $\Gamma_0\ll\gamma$ due to total internal reflection at the surfaces of the patterned slab  \cite{Arcari2014}. Thus, for now we neglect coupling to free-space modes. (The effect of coupling to free-space photons is discussed in detail in Ref.~\cite{SM_ref}.)

The single excitation eigenmodes of \refeq{Hamiltonian} are Bloch modes of the form
\bal\label{BlochModes}
\ket{\psi_{\bk}}=\sum\limits_n e^{\text{i}\bk\cdot \bR_n}\Big[c_{+,\bk}\ket{\sigma_{+,n}}+c_{-,\bk}\ket{\sigma_{-,n}}\Big],\quad\;
\eal
where the summation runs over all lattice vectors $\{\bR_n\}$ and $\bk$ is the Bloch quasi-momentum. For each $\bk$, there are two eigenvalues $\omega_{\bk}$ that can be numerically calculated from the photonic band structure shown in \reffig{fig:atom_bands}(a) \cite{SM_ref}. \reffig{fig:atom_bands}(b) shows the hybridized atomic-photonic bands in the immediate vicinity of the $\mathbf K$ symmetry point in the presence of an out-of-plane magnetic field. The bare photonic bands of the Dirac cone are also shown schematically as yellow dashed lines for reference. One of the atomic bands does not interact with the guided slab modes due to polarization mismatch, forming a flat band in the middle. The other atomic band hybridizes with the guided modes, forming avoided crossings that split the band into two disjoint parts. Two equal band gaps form just above and below the middle band and the three bands have Chern numbers -1, +2 and -1, respectively.

\begin{figure}[h!]
\centering
\includegraphics[width=.475 \textwidth]{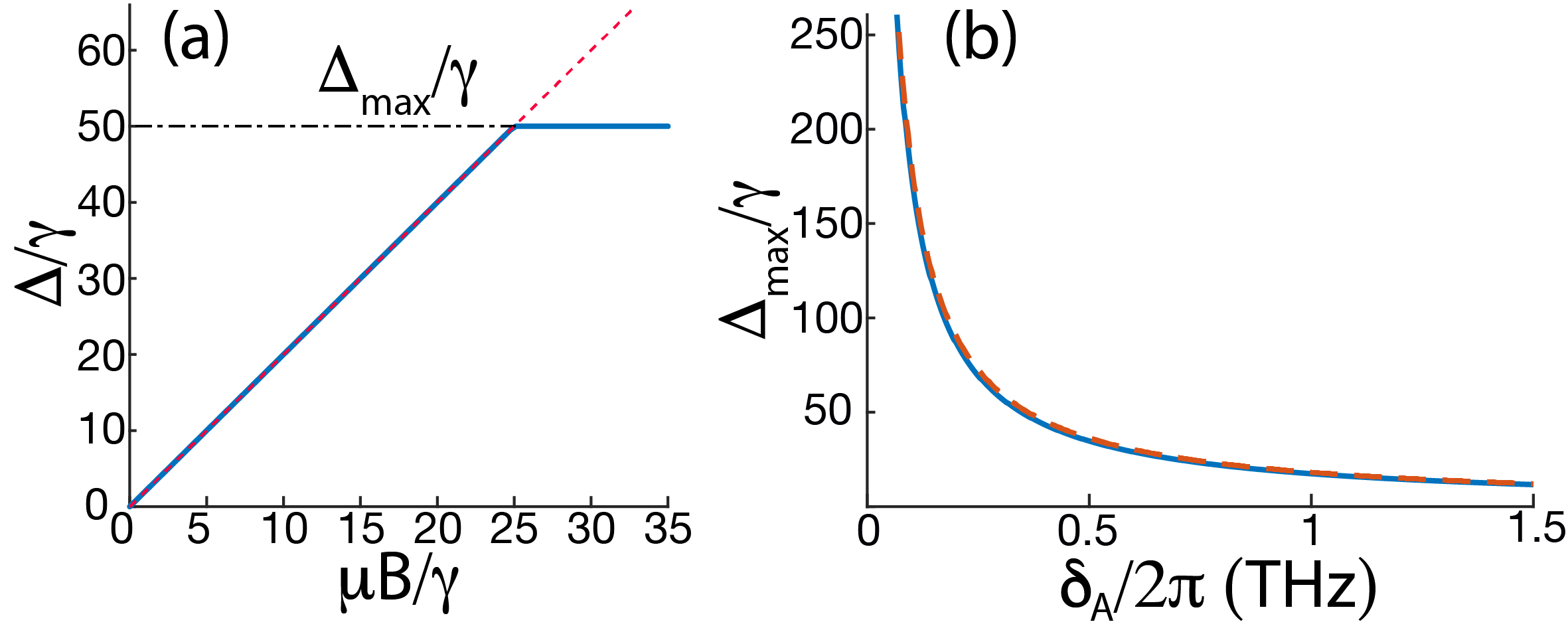}
\caption{
\label{fig:gap_scaling}
(a)  Gap size scales linearly with B-field ($\Delta = 2\mu B$) until a plateau is reached at $\Delta_\text{max}$ (blue line). (b) Maximum gap size $\Delta_\text{max}$ as a function of the detuning $\delta_A=\omega_\text{Dirac}-\omega_A$ (blue line). The dashed red line is a $\sim1/\delta_A$ fit.} 
\end{figure} 

The size of the band gap $\Delta$ between topological bands is of particular importance, since the topological protection of the system (e.g. the robustness of edge states to defects) increases with gap size. Since the gap arises from the Zeeman-splitting of the atomic levels \cite{Perczel2017a}, the gap size is a linear function of the applied magnetic field (\reffig{fig:gap_scaling}(a)), until a plateau is reached at $\Delta_\text{max}$ when the middle band completely flattens. The maximum achievable gap size $\Delta_\text{max}$ is plotted in \reffig{fig:gap_scaling}(b) as a function of $\delta_A$. As the tip of the photonic Dirac cone is tuned closer to the atomic frequency, the energy gap increases as $\sim 1/\delta_A$. Ultimately, the gap size is limited by the fact that our quantum optical model (which relies on the accuracy of the Markov approximation near the Dirac cone \cite{Gonzalez-Tudela2018}) is only valid as long as the emitter-field correlation time $\tau_c$ is much shorter than the typical timescale on which the atomic system evolves $\tau_A$ (see Ref.~\cite{SM_ref} for more details).

\begin{figure}[h!]
\centering
\includegraphics[width=.475 \textwidth]{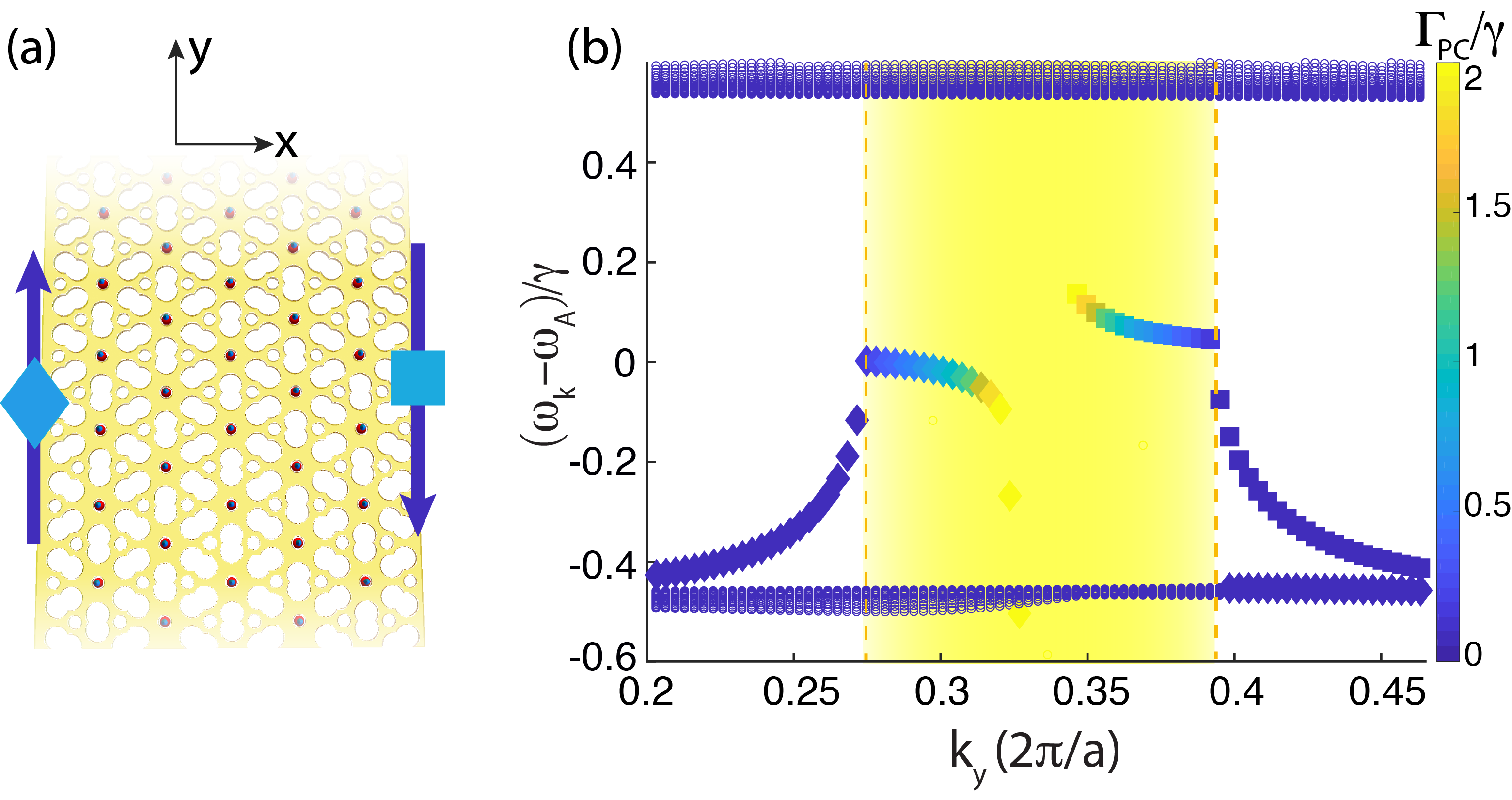}
\caption{
\label{fig:edge_states}
(a) An infinite stripe of emitters embedded in an extended photonic crystal slab. (b) Edge states inside the band gap near the $\mbf K$ point. Unidirectional edge states on the left (right) boundary are indicated by diamonds (squares). The bulk modes are marked by dots, forming the two nearly flat bands near the top and bottom. Yellow shading and yellow dashed lines mark the region where bands can decay into resonant slab modes. Color coding shows the decay rate $\Gamma_\text{PC}$. Relevant parameters are $\delta_A$=18.73THz, $v_\text{s}=0.25c$, $|E_0|^2=0.1855/a^3$, $\Delta = \gamma$ and $\mu B = 0.5\gamma$. Results were obtained for a stripe with 41 atoms in the transverse direction. States that have 5 times more overall amplitude on the five leftmost emitter columns than on the five rightmost columns, are classified as edge states.} 
\end{figure} 

We next explore the topological edge states of the system. We consider a stripe of emitters, embedded in an extended PhC, that is infinite along the $y$-axis, but finite along the $ x$-axis as shown in Fig.~\ref{fig:edge_states}(a). 
Fig.~\ref{fig:edge_states}(b) shows the unidirectional edge states inside the gap near the $\mbf K$ point \cite{footnote_PC_1}. These states are confined to the system boundaries and carry energy only in a single direction, reflecting the broken time-reversal symmetry and the topological protection of the system \cite{Laughlin1981,Halperin1982,Haldane2008,Raghu2008,Wang2008,Perczel2017a,Perczel2017b}. Edge states on the left side of the stripe are marked with triangles, whereas edge states on the right side are marked with squares. The quasi-momentum region where the guided modes of the Dirac cone are resonant with the emitters is shaded yellow and its boundaries are marked by yellow dashed lines (see yellow dashed lines in \reffig{fig:atom_bands}(b) for reference). In this region the emitters can directly couple to guided slab modes with decay rate $\Gamma_\text{PC}$, the magnitude of which is indicated in the figure by the color code. Modes that fall outside the yellow region cannot couple to slab modes due to the momentum mismatch, 
making these modes long-lived. Note that a different edge termination with slightly different edge state properties is obtained when the triangular lattice is rotated $\pi/6$ relative to the one shown in Fig.~\ref{fig:edge_states}(a) (see Ref.~\cite{SM_ref} for more details).

\begin{figure}[h!]
\centering
\includegraphics[width=.45 \textwidth]{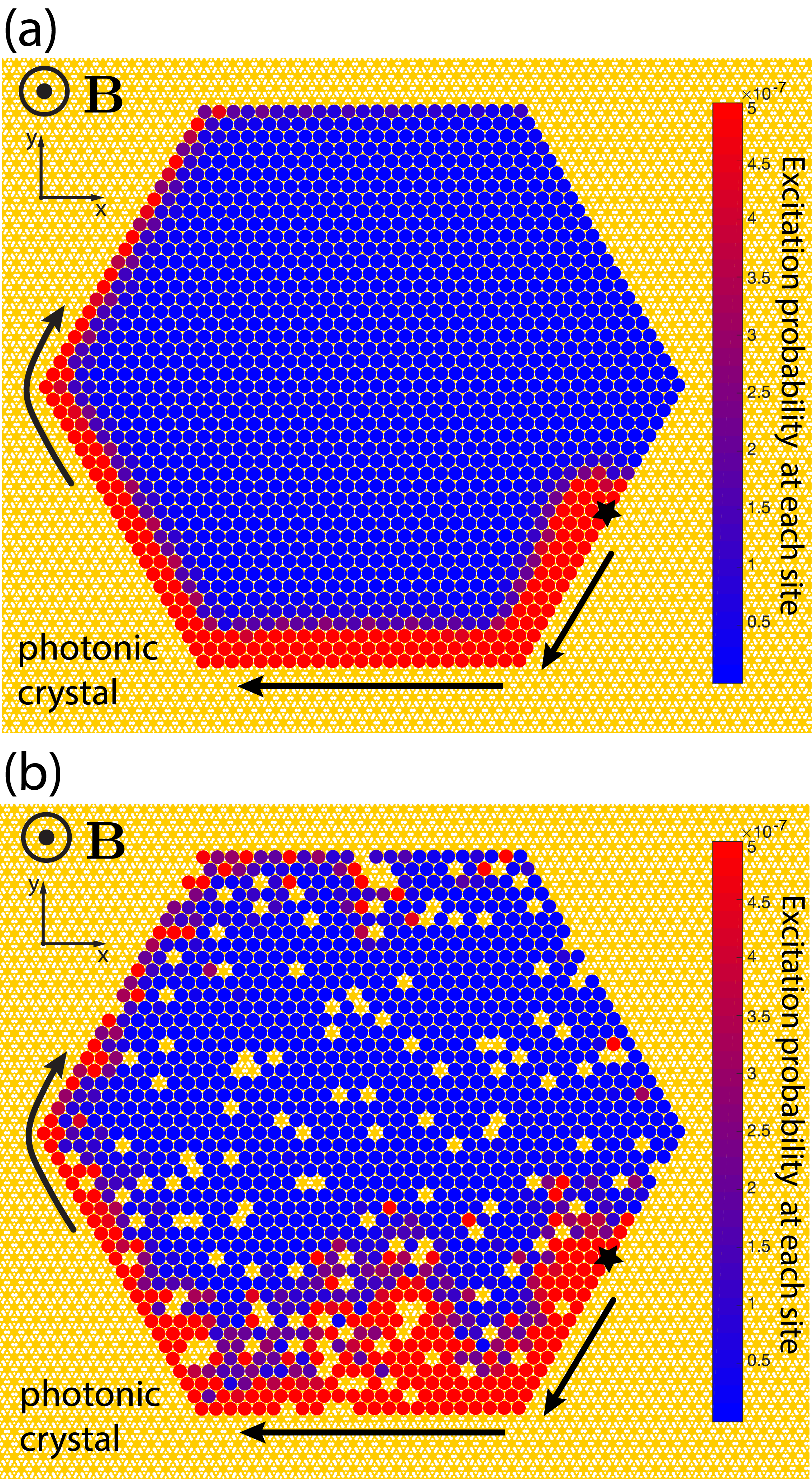}
\caption{
\label{fig:real_space}
(a) Time-evolved state ($t=161.5\gamma^{-1}$) of a hexagonal-shaped triangular lattice of emitters ($N=1519$) embedded in an extended photonic crystal slab. An emitter at the edge is addressed by a laser (black star). The edge state propagates on the boundaries without reflection or significant decay into slab modes. The color code shows the excitation probability at each emitter. (b) Time-evolved state ($t=297.5\gamma^{-1}$) state of the system ($N=1380$) in the presence of imperfect lattice filling (90\%) and inhomogeneous broadening, where the emitter frequencies are sampled from
${P(\omega)=1/\sqrt{2\pi \sigma^2}\exp(-(\omega-\omega_A)^2/(2\sigma^2)}$, where $\sigma=0.1\Delta$. Other relevant parameters are $\delta_A/2\pi = 18.73$THz, $v_\text{s}=0.25c$, $|E_0|^2=0.1855/a^3$, $\Delta = \gamma$, $\mu B = 0.5\gamma$, $\Omega = 0.0059\gamma$ and $\omega_L = \omega_A - 0.37\gamma$. The laser drive is switched on adiabatically with the profile ${\Omega(t)=\Omega \exp(-[t-t_0]^2/[2\Sigma^2])}$, where $t_0 =127.5\gamma^{-1}$ and $\Sigma = 23.3\gamma^{-1}$.} 
\end{figure} 

Next, we study the time evolution of edge states in real space. \reffig{fig:real_space}(a) shows a defect-free triangular lattice of emitters with a hexagonal shape. The emitters are embedded in an extended PhC, whose spatial dimensions are much larger than those of the atomic lattice. We assume that one emitter on the edge is addressed by a weak laser with frequency $\omega_L$ that is resonant with the edge states in the lower half of the gap. The laser couples to the $\sigma_+$ and $\sigma_-$ transitions with equal strength $\Omega$, such that $\Omega\ll \Delta$. We calculate the time evolution of the wavefunction of the initial state by numerically exponentiating \refeq{Hamiltonian} \cite{Perczel2017b}. The evolved state of the system at one particular time is shown in \reffig{fig:real_space}(a), where the color of each site reflects the excitation probability of the emitter. The edge states excited by the laser carry energy only in the clock-wise direction, reflecting the broken time-reversal symmetry of the system. Coupling to bulk modes is strongly suppressed. Furthermore, the excitation routes around the lattice corners without reflection or significant loss into the guided modes of the  slab.   

The hallmark of topological edge states is their robustness to significant imperfections in the system. \reffig{fig:real_space}(b) shows a lattice, where a random $10\%$ of the emitters is missing. Furthermore, we assume that the inhomogeneities in the nanophotonic environment (e.g. fabrication imperfections) give rise to the inhomogeneous broadening of emitters. Therefore, we sample the emitter frequencies from a Gaussian distribution, whose width is 10\% of the gap. \reffig{fig:real_space}(b) shows a snapshot of the time-evolved state of the system. Despite the presence of the lattice defects and the inhomogeneity of emitters, the edge state propagates along the system edges, routing around all defects, including the cluster of missing emitters near the bottom edge. These results demonstrate the topological robustness of the edge states and the system as a whole.

The flat middle band in \reffig{fig:atom_bands}(b) is another key feature of our topological system. The flatness of the band and the non-trivial Chern number are, in general, the two key ingredients for fractionalized topological phases in lattice models \cite{Neupert2011,Tang2011,Sun2011,Peter2015,Yao2012,Yao2013,Wang2012,Yang2012,Trescher2012}. Usually,  such flat bands require careful tuning of the long-range hopping parameters. Here, away from the $\mbf K$ point the band remains flat due to the fact that the guided photonic bands are far-detuned and coupling to them is weak (\reffig{fig:atom_bands}(a)). Near the $\mbf K$ point, flatness arises from the combined effects of the uniform magnetic field and the fact that the middle band does not couple to the PhC bands due to polarization mismatch. 
The inherent non-linearity of the constituent emitters gives rise to a hard-core constraint for the bosonic excitations of the system, which provides a point-like interaction for the excitations. Such short-range interactions are ideal for the realization of interacting many-body topological phases \cite{Gong2016}. Within such a system 
we expect non-trivial competition between states with strongly correlated topological phases and those that behave analogously to electronic charge density waves. This competition arises from the fact that the atoms interact primarily through the Dirac cone and, therefore, the Berry curvature is concentrated in the vicinity of the $\mbf K$ points. Thus only states near the $\mbf K$ points are expected to yield exotic topological many-body phases, whereas the rest of the band will likely contribute to states that resemble an ordered quantum fluid. Controlling the distribution of the Berry curvature in the flat band (e.g. using established techniques from cold atom experiments \cite{He2017}) can be explored for engineering experimentally observable fractional topological phases. This will be addressed in forthcoming work.

While practical realization of the proposed systems constitutes a non-trivial challenge, most  of the key ingredients have already been demonstrated recently.  For example, a hybrid system based on 
SiV color defects in diamond can be created by first fabricating a thin diamond slab \cite{Faraon2012,Wan2018}, SiVs can be implanted using focused ion beam implantation \cite{Pacheco2017}, while GaP can be positioned on both sides of the slab \cite{Aparna2014} with the air-hole structure created by etching through the GaP and diamond layers. Alternatively, a PhC hole structure made entirely of diamond can potentially be used (see Ref.~\cite{SM_ref} for more details).

We have demonstrated that two-dimensional quantum emitter arrays embedded in photonic crystals constitute a topological quantum optical system. The atomic bands have non-trivial Chern numbers in the presence of an out-of-plane magnetic field and the dispersion of one of the bands is significantly quenched. Long-lived topological edge states appear on the system boundaries, which are robust to imperfect lattice filling as well as to inhomogeneous broadening of the emitters. We believe that the experimental accessibility of this platform will open up exciting opportunities for exploring novel topological states of light and matter, including quantum optical analogues of fractional Chern insulators.

We would like to thank Mihir Bhaskar, Ruffin Evans, Alejandro Gonz\'alez-Tudela, Fabian Grusdt, Bert Halperin, Hannes Pichler, Denis Sukachev, Efraim Shahmoon, Dominik Wild, Bihui Zhu and Peter Zoller for valuable discussions. This  work  was  supported through the National Science Foundation (NSF), the MIT-Harvard Center for Ultracold Atoms, the Air Force Office of Scientific  Research  via  the  MURI,  the  Vannevar  Bush Faculty Fellowship and DOE. Some of the computations in this paper were performed on the Odyssey cluster supported by  the  FAS  Division  of  Science,  Research  Computing Group at Harvard University. J. P. acknowledges support from the Dr. Elem\'er and \'Eva Kiss Scholarship Fund. J. B. acknowledges support from the European Research Council (ERC Grant Agreement no. 337603) and VILLUM FONDEN via the QMATH Centre of Excellence (Grant no. 10059). D. E. C. acknowledges support from the ERC Starting Grant FOQAL, MINECO Plan Nacional Grant CANS, MINECO Severo Ochoa Grant No. SEV-2015-0522, CERCA Programme/Generalitat de Catalunya, Fundacio Privada Cellex and AGAUR Grant 2017 SGR 1334.

\clearpage

\pagebreak
\onecolumngrid
\begin{center}
\textbf{\large Supplemental Material}
\end{center}
\setcounter{equation}{0}
\setcounter{figure}{0}
\setcounter{table}{0}
\setcounter{page}{1}
\makeatletter
\renewcommand{\theequation}{S\arabic{equation}}
\renewcommand{\thefigure}{S\arabic{figure}}
\renewcommand{\bibnumfmt}[1]{[S#1]}
\renewcommand{\citenumfont}[1]{S#1}

\vspace{.5cm}

\twocolumngrid

The Supplemental Material is organized as follows. In Sec.~1, we discuss the calculations for obtaining the atomic band structure. In Sec.~2, we describe the real-space Green's function and its properties. 
In Sec.~3, we discuss the validity of our model and the limits on the maximum achievable gap size in our system. In Sec.~4, we describe the calculations behind obtaining the edge states in our system. Finally, in Sec.~5, we analyze the edge states in the presence of free space decay.
 
\section{Calculation of the atomic band structure} \label{Section1}

In this section, we calculate the Bloch modes of the hybrid emitter-photonic crystal system. For simplicity, we focus on the atomic interactions mediated by the photonic crystal modes and neglect coupling to free-space photons. This is motivated by previous studies of a triangular lattice of emitters in free space \cite{SM_Perczel2017b}. The cooperative effects mediated by free-space photons reported in Ref.~\cite{SM_Perczel2017b} are, in general, an order magnitude smaller than those mediated by the photonic crystal modes described here.

\subsection{Analytic calculations}

The Bloch eigenmodes of our system with quasi-momentum $\bk$ can be obtained by substituting Eqs.~(1) and (4) from the Main Text into ${H\ket{\psi_\bk}=\hbar \omega_{\bk}\ket{\psi_\bk}}$. After transforming to a Cartesian basis using the relation ${\ket{\sigma_\pm}=\mp (\ket{x}\pm i\ket{y})/\sqrt{2}}$, the Bloch modes are obtained by diagonalizing the following 2x2 matrix 
\bal\label{Matrix2x2}
M_{\alpha\beta}(\bk)&=&\left(\tilde\omega_A - \text{i}\frac{\Gamma_\text{PC}}{2}\right)\delta_{\alpha\beta}  + \xi_{\alpha,\beta}\nonumber\\
&+& \fr{3\pi\gamma  c}{\omega_An_\text{d}}\sum\limits_{\bR\neq 0}e^{\text{i}\bk\cdot \bR}G_{\alpha\beta}(\bR),
\eal
where $\alpha,\beta =x,y$ label the polarization components and $\tilde\omega_A$ denotes the `dressed' atomic transition frequency in the photonic crystal. The summation is implied over the set of triangular lattice vectors $\{\bR = n_1\bR_1+n_2\bR_2\}$, where $n_1,n_2=0,\pm 1, \pm 2,\dots$ and
\bal
\bR_1 = a\left(\fr{\sqrt{3}}{2},\fr{1}{2}\right)\quad\text{and}\quad \bR_2 = a\left(\fr{\sqrt{3}}{2},-\fr{1}{2}\right).
\eal
The Zeeman splitting of the atomic levels is given by
\bal
\xi_{\alpha,\beta} = - \text{i}\mu B(\delta_{\alpha x}\delta_{\beta y}-\delta_{\alpha y}\delta_{\beta x}).
\eal
To simplify \refeq{Matrix2x2}, we write the atomic transition frequency as 
\bal
\tilde\omega_A =  \omega_A + \delta\omega_\text{PC},
\eal
where $\omega_A$ is the `bare' atomic transition in bulk diamond and $\delta\omega_\text{PC}$ is the energy shift of the individual atoms due to the presence of the photonic crystal environment. We make use of the fact that 
\bal\label{GreensExp}
 \fr{3\pi\gamma \hbar c}{\omega_An_\text{d}}G_{\alpha\beta}(\mbf 0) = \delta\omega_\text{PC} - \text{i}\frac{\Gamma_\text{PC}}{2}
\eal
to rewrite \refeq{Matrix2x2} as
\bal\label{Matrix2x2Simple}
M_{\alpha\beta}(\bk)=\omega_A\;\delta_{\alpha\beta}+\xi_{\alpha,\beta}+  \fr{3\pi\gamma  c}{\omega_An_\text{d}}\sum\limits_{\bR}e^{\text{i}\bk\cdot \bR}G_{\alpha\beta}(\bR).\qquad
\eal
Generally, it is difficult to work with the Green's function in real space, due to the slow convergence of the sum above. Therefore, it is convenient to transform the summation over real-space lattice vectors $\{\bR\}$ to a summation over momentum-space reciprocal lattice vectors $\{\bG=n_1\bG_1+n_2\bG_2\}$, where
\bal
\bG_1 = \fr{2\pi}{a}\left(\fr{1}{\sqrt{3}},1\right)\quad\text{and}\quad  \fr{2\pi}{a}\left(\fr{1}{\sqrt{3}},-1\right).
\eal
The momentum-space summation can be performed over the Fourier-transform of the Green's function in momentum space. 

In particular, we substitute into \refeq{Matrix2x2Simple} the following expression
\bal
G_{\alpha\beta}(\br)=\int_{BZ}\fr{d^2\bp}{(2\pi)^2}g_{\alpha\beta}(\bp)e^{\text{i}\bp\cdot \br},
\eal
where the integral is performed over the irreducible Brillouin zone of the photonic crystal and $g_{\alpha\beta}(\bp)$ stands for the Green's function in momentum space.
Next, we make use of the following form of Poisson's identity
\bal
{\sum\limits_{\bR}e^{\text{i}(\bk+\bp)\cdot\bR}}={\fr{1}{\mathcal{A}}\sum\limits_\bG (2\pi)^2\delta^{(2)}(\bp+\bk-\bG)}
\eal
to transform the summation to momentum space, where $\mathcal A$ is the area of the hexagonal unit cell in real space. We perform the integral to obtain 
\bal
&&\fr{1}{\mathcal{A}}\sum\limits_{\bG}\int_{BZ}d^2\bp \; g_{\alpha\beta}(\bp)\delta^{(2)}(\bp-(\bG-\bk))\qquad\nonumber\\
&&\quad= \fr{1}{\mathcal{A}}g_{\alpha\beta}(-\bk)=\fr{1}{\mathcal{A}}g_{\alpha\beta}(\bk),
\eal
where the first equality follows from the fact that only the Dirac delta with $\bG=0$ contributes to the integral (since the integral is restricted to the irreducible Brillouin zone), and the second equality follows from time-reversal symmetry of the photonic crystal spectrum \cite{SM_Bernevig2013}. Using this expression, we can rewrite \refeq{Matrix2x2Simple} as 
\be\label{Matrix2x2Momentum}
M_{\alpha\beta}(\bk)=\omega_A\;\delta_{\alpha\beta} +\xi_{\alpha,\beta}+  \fr{3\pi\gamma  c}{\omega_An_\text{d}}\fr{1}{\mathcal{A}}g_{\alpha\beta}(\bk).
\ee

Finally, we expess the momentum-space Green's function in terms of its eigenvalue expansion \cite{SM_Perczel2018} as 
\be\label{GreensMomentum}
g_{\alpha\beta}(\bk) =\mathcal{A}c^2 \sum\limits_n\fr{E_{\bk,\alpha}^{(n)*}(\br_A)\,E_{\bk,\beta}^{(n)}(\br_A)}{\omega_A^2-\big(\omega_\bk^{(n)}\big)^2},
\ee
where $E^{(n)}_{\bk,\alpha}(\br)$ denotes the $\alpha$ component ($\alpha=x,y$) of the electric field of the photonic crystal mode in the $n^\text{th}$ band at quasi-momentum $\bk$, while $\omega_\bk^{(n)}$ is the corresponding frequency of the mode. 


 Even though $E^{(n)}_{\bk,\alpha}(\br)$ gives the electric field distribution inside the entire real-space unit cell, we only sample the field at the position of the emitter ($\br=\br_A$), which we choose to be at the geometric center of the hexagonal cell and $z=0$ to ensure that the emitter only couples to TE-like modes \cite{SM_Joannopoulos2008}. 
 
Due to the underlying periodicity of the photonic crystal lattice, the electric field mode can be expressed in a canonical Bloch form as 
\bal\label{normalizedU}
E^{(n)}_{\bk,\alpha}(\br) = \fr{u^{(n)}_{\bk,\alpha}(\br)}{\sqrt{a^3}}e^{\text{i}\bk\cdot \br},
\eal
where $a=|\bR|$ is the periodicity of the lattice and ${\mbf u^{(n)}_\bk(\br+\bR)=\mbf u^{(n)}_\bk(\br)}$ is a dimensionless periodic vector function that is normalized to ensure that  
\bal
\int_\mathcal{V}d^3\br\; \vare(\br)\,  \mbf{E}_{\bk}^{(n)}(\br)\cdot \mbf{E}_{\bk'}^{(n)*}(\br) = \delta_{\bk\bk'},
\eal
where $\vare(\br+\bR)=\vare(\br)$ is the periodic dielectric permittivity function describing the photonic crystal in real space and the integral is performed over the quantization volume \cite{SM_Glauber1991,SM_Perczel2018}.

\subsection{Numerical calculations using MPB}

We use the open-source MIT Photonic Bands 1.4.2 (MPB) numerical software package \cite{SM_Johnson2001} to directly obtain the expressions for $\omega_\bk^{(n)}$ and the normalized $u^{(n)}_{\bk,\alpha}(\br_A)$ \footnote{S. G. Johnson, Official MPB Documentation (online), https://mpb.readthedocs.io/en/latest/\\
Scheme\_User\_Interface/\#field-normalization}. MPB is an iterative eigensolver that uses a planewave basis to iteratively improve approximations to the eigenstates and eigenvalues of Maxwell's equations. MPB approximates the solution using a planewave cutoff, which corresponds to the spatial discretization of the unit cell. In our simulations, we take a supercell of height 4a and run the calculations with a supercell discretization resolution of $256\times 256\times 64$.

In particular, we perform the numerical calculations by defining a rhombic 2D supercell that contains a $3\times 3$ triangular array of holes with hole spacing $d=a/3$. We remove the central hole and push the surrounding 6 holes radially outward (\reffig{fig:unit_cell}), such that the center of these 6 holes is a distance $1.2d$ away from the center. The radii of these 6 holes is increased to $r_\text{l}=0.35d$. We leave the other two holes in place, which have a radius of $r_\text{s}=0.25d$. The slab is composed of three different layers. In the middle, we define a diamond ($\vare_\text{d}=5.76$) layer of thickness $0.5a$, which is sandwiched between two GaP ($\vare_\text{GaP}=10.5625$) layers of thickness $0.315a$. In the resulting band structure, the tip of the Dirac cone is at $\omega a/(2\pi c)=0.32545$. The eigenenergies $\omega_\bk^{(n)}$ and the corresponding field intensities $|u^{(n)}_{\bk,\alpha}(\br_A)|^2$ of the TE-like modes at the location of the emitters ($\br_A=\mbf 0$) are plotted in Fig.~2(a) of the Main Text. After substituting $\omega_\bk^{(n)}$ and $u^{(n)}_{\bk,\alpha}(\br_A)$ into \refeq{normalizedU} and \refeq{Matrix2x2Momentum}, we diagonalize the $2\times 2$ matrix in \refeq{GreensMomentum} and we obtain two eigenvalues $\omega_\bk^{(m)}$ (${m=1,2}$) for each $\bk$. The results are plotted in Fig.~2(b) of the Main Text for a non-zero magnetic field.

\subsection{All-diamond photonic crystal slab}

\begin{figure}[h!]
\centering
\includegraphics[width=0.45 \textwidth]{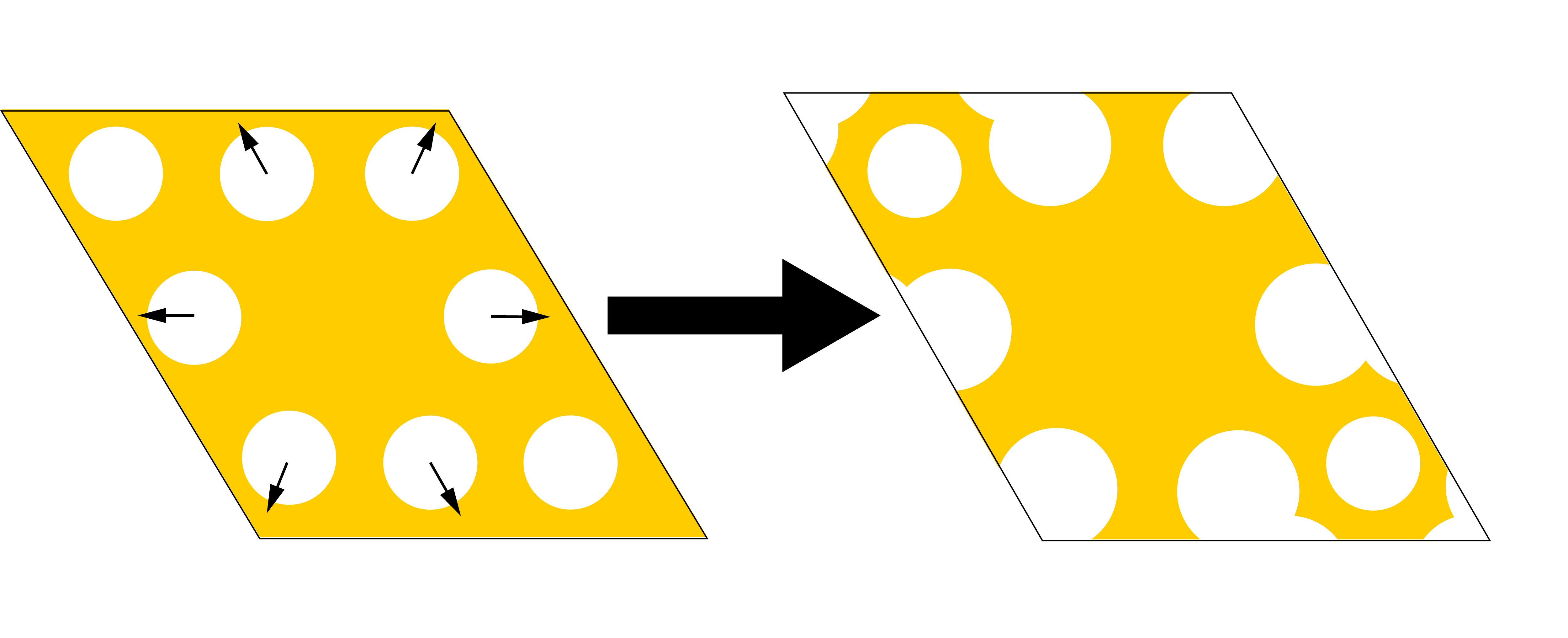}
\caption{
\label{fig:unit_cell}
Constructing the unit cell for the photonic crystal. We take a rhombic supercell with a $3\times 3$ triangular array of holes with spacing $d$ and radius $0.25d$. The central hole is removed and the surrounding 6 holes are pushed radially outward to a distance of $1.2d$, while also increasing their radii to $0.35d$. The spacing between units cells is $a=3d$.} 
\end{figure}

Note that it is also possible to use an all-diamond photonic crystal with SiVs, which would significantly simply the fabrication process. An all-diamond structure with the cavity-like arrangement of air holes shown in \reffig{fig:unit_cell} would also have a stand-alone Dirac cone and give rise to dipolar interactions with winding phases \cite{SM_Perczeletal2018}. However, given that diamond has a lower refractive index than GaP, tuning the photonic Dirac cone resonant with the SiVs would require a photonic structure with significantly larger lattice spacing $a$ than for the hybrid structure. This reduces the size of the irreducible Brillouin zone in $k$-space to the extent that the entire Brillouin zone falls within the light cone region ($k<2\pi/\lambda$). This, in turn, implies that all of the edge states of the system could couple to free-space photons (see Sec.~5. for more details).

\section{The real-space Green's function} \label{SectionNew}

In order to calculate the time evolution of a finite lattice of emitters (e.g. Fig.~5 in the Main Text), we need to directly substitute the real-space Green's function into Eq.~(1) of the Main Text. The explicit derivation of the real-space Green's function of our system is discussed in detail in Ref.~\cite{SM_Perczeletal2018}. Here we simply give its explicit form and briefly describe its properties. 

The dipolar coupling mediated by the photonic Dirac cone is described by the following Green's function \cite{SM_Perczeletal2018}
\bal\label{Greens}
\mbf G(\br)\!=\!	\left[
		\begin{array}{cc}
		-P^+(\br)H_0^{(2)}(r/\xi ) & e^{\text{i}\phi}P^-(\br)H_1^{(2)}(r/\xi)\\[6pt]
		-e^{-\text{i}\phi}P^-(\br)H_1^{(2)}(r/\xi ) & -P^+(\br)H_0^{(2)}(r/\xi )
		\end{array}
	\!\right]\!\!,\qquad\;
\eal 
where $\br\!=\!r(\cos\phi,\sin\phi)$ is the position vector, the diagonal (off-diagonal) terms of $\mbf G$ account for the coupling between atomic transitions of the same (opposite) circular polarization, $H_m^{(2)}$ denotes the Hankel function of the second kind of order $m$ and $\xi = v_\text{s}/\delta_A$ gives the length scale of the interaction as a function of the linear dispersion of the slab modes at the Dirac cone $v_\text{s}$ and the atomic detuning from the tip of the Dirac cone $\delta_A=\omega_\text{Dirac}-\omega_A$. The prefactors are given by 
\bal
P^{\pm}(\br) = \text{i}\fr{\mathcal Ac^2|E_0|^2\delta_A}{8\omega_Av_\text{s}^2}\left(  e^{\text{i}\bp_\mathbf{K}\cdot \br} \pm  e^{\text{i}\bp_{\mathbf{K}'}\cdot \br} \right),\eal
where $\mathcal A=\sqrt{3}/2a^2$ is the area of the hexagonal unit cell, $|E_0|^2=|\mbf E_\bk(\br_A)|^2$ is the electric field intensity in the vicinity of the Dirac cone, which is approximately $\mbf k$-independent, and $\bp_\mathbf{K}$ and $\bp_{\mathbf{K}'}$ are the quasi-momenta associated with the two inequivalent $\mathbf K$ points in the Brillouin zone. The parameters $v_\text{s}$ and $E_0$ are obtained numerically \cite{SM_Perczeletal2018}. This analytic Green's function quantitatively captures the slab-mediated dipolar interaction, when $\delta_A$ is small. Note the presence of the winding phases $e^{\pm \text{i}\phi}$ \cite{SM_Peter2015,SM_Karzig2015,SM_Bettles2017} in the off-diagonal terms of $\mbf G$, which give rise to the topological properties of the system. These winding phases arise from the propagation-direction-dependent polarization structure of the photonic modes of the Dirac cone \cite{SM_Perczeletal2018}. 

\section{Model validity and maximum gap size}\label{validity}

In Fig.~3(b) of the Main Text we found that the size of the gap is inversely proportional to $\delta_A$, the detuning of the atomic frequency from the tip of the photonic Dirac cone. Eventually, as $\delta_A$ is decreased, our theory breaks down, limiting the maximum achievable gap size. In this section we discuss for what parameters this breakdown occurs.

%
%

The non-Hermitian Hamiltonian in Eq.~(1) of the Main Text describes the 
 evolution of the system in the absence of quantum jumps and is obtained from the canonical master equation for open quantum systems \cite{SM_Perczel2017b}. Since the derivation of the master equation assumes the validity of the Born-Markov approximation \cite{SM_Gross1982}, our results are guaranteed to be valid only as long as these approximations hold. However, the Born-Markov approximation is known to break down as the atomic emitters are tuned close to the Dirac vertex \cite{SM_Gonzalez-Tudela2018}.

\begin{figure}[h!]
\centering
\includegraphics[width=0.45 \textwidth]{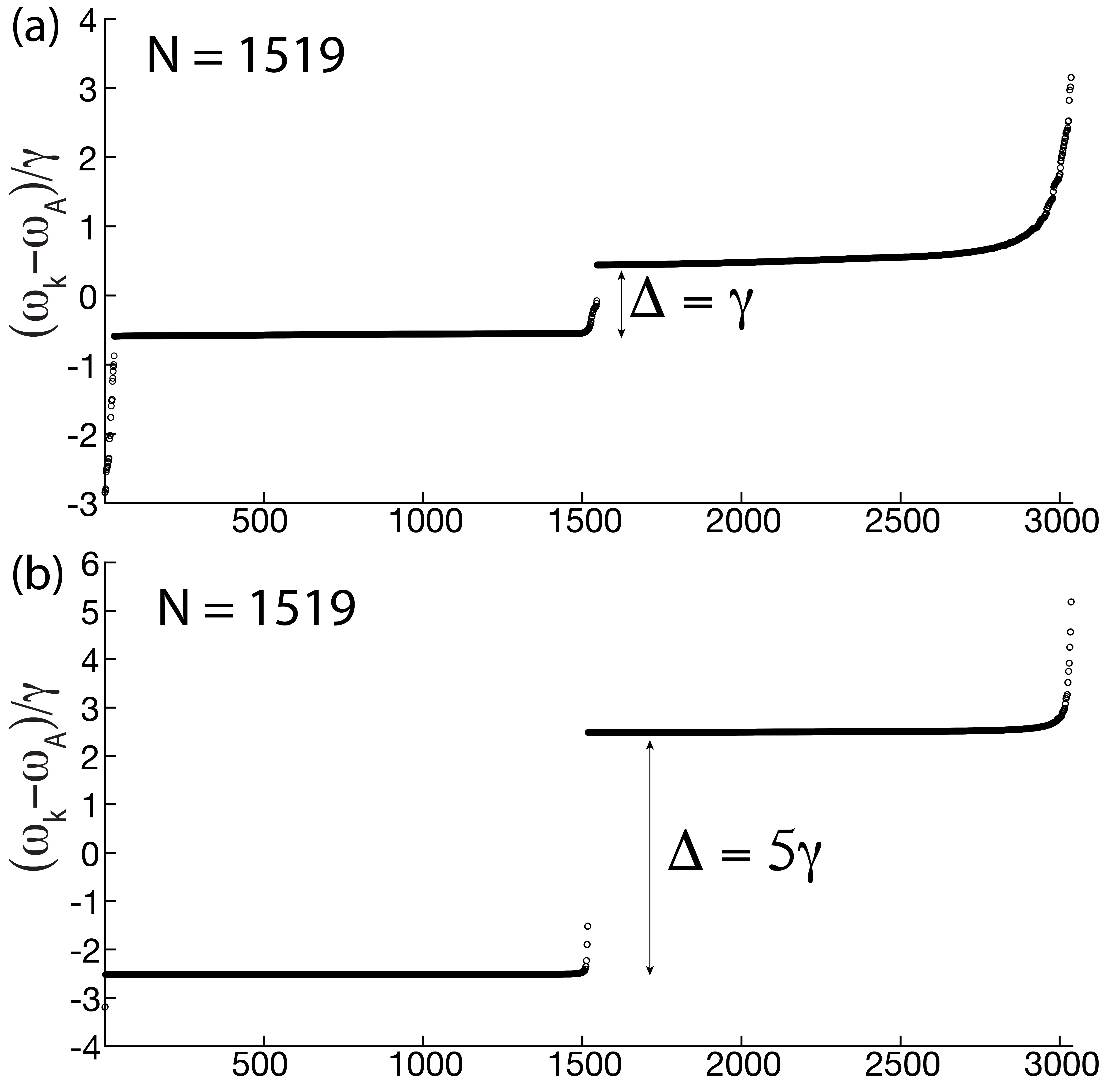}
\caption{
\label{fig:spectra}
(a) Spectrum for a hexagonal array of $N=1519$ emitters with $\delta_A=18.73$THz, $\Delta = \gamma$, $\mu B=0.5\gamma$ and $a=\lambda/3$. Eigenstates are sorted in ascending order. The edge states form a dense set in the bottom half of the gap, leading to the dispersive time-evolution of the edge states shown in Fig.~5(a) of the Main Text. (b) Spectrum for a hexagonal array of $N=1519$ emitters with $\delta_A=3.78$THz, $\Delta = 5\gamma$, $\mu B=2.5\gamma$ and $a=\lambda/3$. The edge states are sparse in the bottom half of the gap due to the reduced density of states for the resonant photonic slab modes. Excitation of the edge states leads to the characteristic few-mode dynamics depicted in \reffig{fig:sparse}. For both (a) and (b), the emitters are assumed to be embedded in the same photonic crystal as in Fig.~1 of the Main Text, with $v_\text{s}=0.25c$ and $|E_0|^2=0.1855/a^3$.  } 
\end{figure} 

The Born approximation assumes that the state of the environment (i.e. the collection of photonic crystal modes) is not significantly affected by its interaction with the emitters. This assumption is guarenteed to hold as long as the photonic crystal is much larger than the emitter array.  

The Markov approximation assumes that the correlation time $\tau_c$ between the environment and the emitters is much smaller than the typical time scale for the evolution of the system $\tau_A$, (i.e. $\tau_c\ll \tau_A$). Here, the correlation time $\tau_c$ of the environment corresponds to the time it takes for a photon to leave the emitter array, since for $t<\tau_c$ the photon can still be reabsorbed by another emitter. Denoting the characteristic size of the emitter array with $L$, we find that $\tau_c\sim L/v_\text{s}$, where $v_\text{s}$ is the group velocity of the guided modes of the photonic crystal slab.

\begin{figure}[h!]
\centering
\includegraphics[width=0.45 \textwidth]{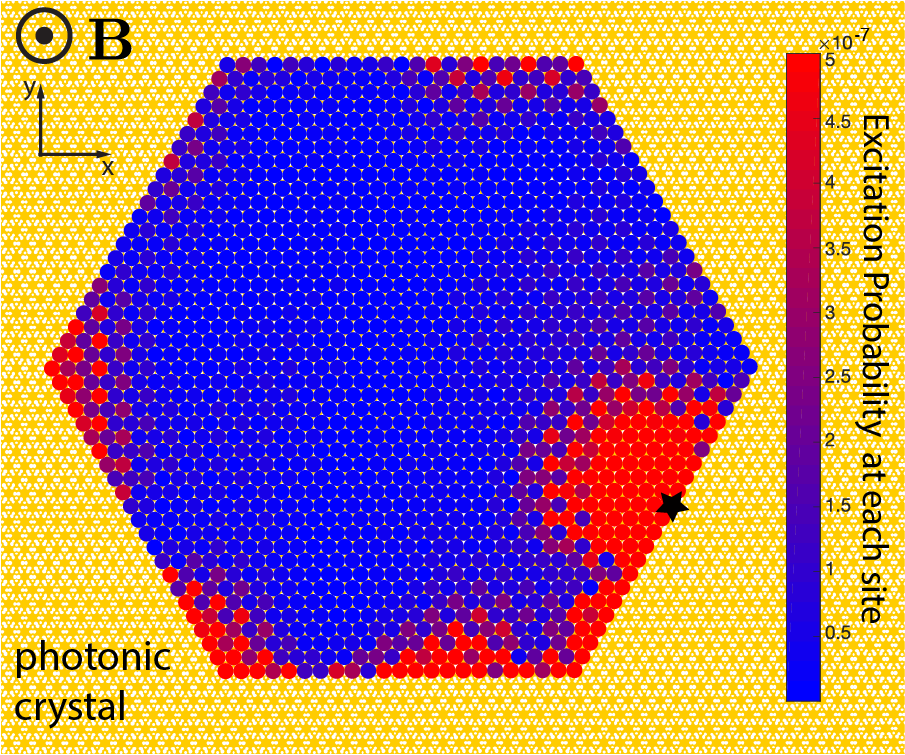}
\caption{
\label{fig:sparse}
(a) Time-evolved state ($t=306\gamma^{-1}$) of a hexagonal-shaped triangular lattice of emitters ($N=1519$), showing the non-dispersive edge state dynamics when the density of photonic states mediating emitter interactions is small. A single emitter at the edge (black star) is addressed by a laser of stength $\Omega$ and frequency $\omega_L$. Relevant parameters are $\delta_A/2\pi = 3.78$THz, $v_\text{s}=0.25c$, $|E_0|^2=0.1855/a^3$, $\Delta = 5\gamma$, $\mu B = 2.5\gamma$, $\Omega = 0.0295\gamma$ and $\omega_L = \omega_A - 1.85\gamma$. The laser drive is switched on adiabatically with the profile ${\Omega(t)=\Omega \exp(-[t-t_0]^2/[2\Sigma^2])}$, where $t_0 =127.5\gamma^{-1}$ and $\Sigma = 23.3\gamma^{-1}$.} 
\end{figure}

Our focus in this work is the excitation and time evolution of edge states, for which the relevant time scale is $\tau_\text{s}\sim\Delta^{-1}$, where $\Delta$ is the energy gap (recall that the group velocity of the edge states is proportional to $\Delta$). Thus the Markov approximation requires that 
\bal\label{condition}
\Delta \ll v_\text{s}/L.
\eal
Note that the system size $L$ and the gap size $\Delta$ cannot be varied independently. The gap size scales as ${\Delta\sim 1/ \delta_A}$, and as $\delta_A$ is reduced, the density of photonic states (DOS) through which the emitters interact decreases as $\text{DOS}\sim \delta_A\sim1/\Delta$. To illustrate how the DOS of the photonic slab modes influences the edges states of our system, in 
\reffig{fig:spectra}(a) and (b) we plot the spectrum for $N=1519$ emitters for $\Delta = \gamma$ ($\delta_A=18.73$THz) and $\Delta = 5\gamma$ ($\delta_A=3.78$THz), respectively. All eigenstates are ordered in ascending order. The energy gap is indicated in both spectra with a double arrow. All eigenstates inside the gap are edge modes. The parameters in \reffig{fig:spectra}(a) correspond to those used for the time-domain simulation in Fig.~5 of the Main Text. Note that the edge states are densely packed in the lower half of the gap. This leads to dispersive edge state propagation on the system boundaries (see Fig.~5 of the Main Text for reference). In contrast, \reffig{fig:spectra}(b) shows the spectrum when $\delta_A$ is reduced by a factor a 5, leading to a 5 times larger band gap, but also to 5 times fewer edge states in the gap. \reffig{fig:sparse}  shows a snapshot from the time-dynamics when these edge states are excited. Given that only a few modes are excited, no dispersive propagation is observed. Instead, patches of delocalized excitations are formed that appear and disappear as the system evolves in time. This interference effect is the expected behavior for the time evolution of a limited set of modes that have similar energies.

In order to keep the number of edge states inside the gap constant while $\Delta$ is increased, the number of emitters has to be increased as $N\sim \Delta$. In general, the required number of emitters for dispersive edge state propagation scales approximately as $N\approx\Lambda n$, where we introduced $n$ to parameterize the size of the gap as $\Delta= n\gamma$ and $\Lambda\approx1600$ is a phenomenological constant that we deduce from \reffig{fig:spectra}(a), where $N\approx 1600 \Delta/\gamma$. Since the size of the emitter array scales as $L\sim\sqrt{N}a$, we can rewrite \refeq{condition} as 
\bal\label{analyticLimit}
n\gamma \ll v_\text{s}/(\sqrt{\Lambda n}a),
\eal 
which, after rearrangement, yields
\bal
n\ll \left(\fr{v_\text{s}}{a\gamma \sqrt{\Lambda}} \right)^{2/3}.
\eal
Substituting $v_\text{s}=0.25c$, $\gamma/2\pi=300$THz, $a=\lambda/3$, $\lambda = 738$nm and $\Lambda\approx1600$, we find that in our system the Markov approximation holds as long as the gap size satisfies
\bal
\Delta \ll 250 \gamma. 
\eal

\section{Calculation of edge states}\label{edgestates}

In this section we describe how to numerically calculate the edge states of the system and analyze them in detail.

\begin{figure}[h!]
\centering
\includegraphics[width=0.475 \textwidth]{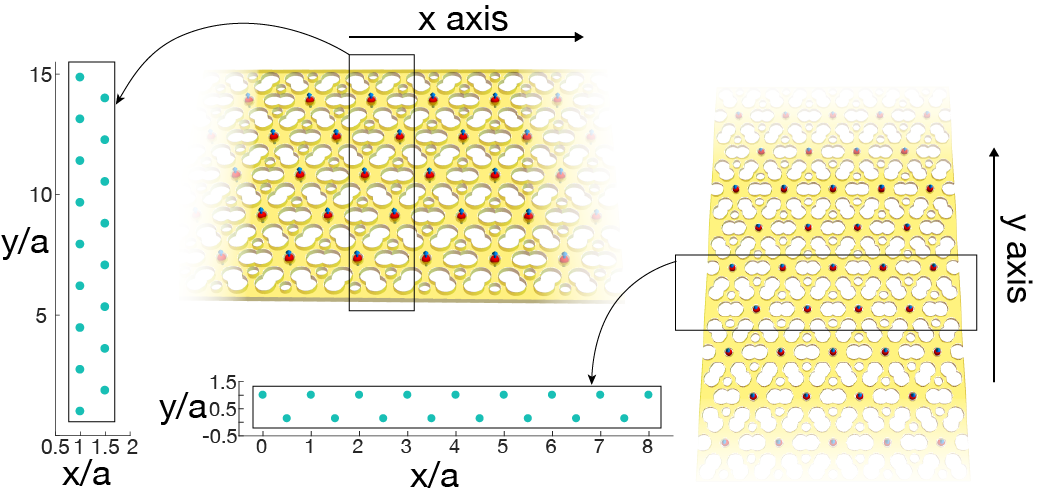}
\caption{
\label{fig:edge_states_calc}
Unit cells for stripes of emitters with two different orientations relative to the triangular lattice. The set of unit cells form 1D lattices with lattice spacing $R_x=a$ along the $x$ axis and $R_y=\sqrt{3}a$ along the $y$ axis. } 
\end{figure}

 We consider stripes of atoms that are infinite along one direction and finite in the other as shown schematically in \reffig{fig:edge_states_calc}. The periodic unit cells of the stripes are identified with black rectangles. The unit cells for both orientations are also shown with $m=17$ atoms. The set of unit cells form a periodic 1D lattice. For the stripe along the x axis the 1D lattice vector is $\mbf R_x=a\hat x$, whereas for the the $y$ axis it is $\mbf R_y = \sqrt{3}a\hat y$. Here, we describe the calculation of the edge states for the stripe oriented along the $y$ axis. The calculation for the other stripe orientation proceeds analogously.

For the stripe oriented along the $y$ axis, the Bloch ansatz takes the following form
\bal \label{kyAnsatz}
\ket{\psi_{k_y}} = \sum\limits_{l=0,\pm1,\dots} \sum\limits_{b=1}^m e^{\text{i}k_y l R_y}\left( c_+^b\ket{\sigma_{+,l}^b}+c_{-}^b\ket{\sigma_{-,l}^b} \right),\qquad 
\eal
where the summation over $l$ is implied over all unit cells, which are placed $|\mbf R_y|=R_y=\sqrt{3}a$ apart along the $y$ axis, $k_y$ is the Bloch quasi-momentum and $b$ labels each of the $m$ emitters inside the unit cell. Acting on \refeq{kyAnsatz} with the Hamiltonian (Eq.~(1) of the Main Text), neglecting the term responsible for coupling to free-space photons and making use of \refeq{GreensExp}, we find that the eigenvalues of the stripe can be obtained by diagonalizing the following ${(2\times m)\times (2\times m)}$ Hamiltonian 
\bal\label{EdgeMatrix}
M_{\alpha\mu,\beta\nu} (k_y)&=& \sum\limits_{b=1}^m\omega_A\delta_{\alpha\beta}\delta_{b\mu}\delta_{b\nu} \nonumber\\
&&\quad + \fr{3\pi\gamma  c}{\omega_An_\text{d}}\sum\limits_{b=1}^m\sum\limits_{b'=1}^m\Lambda_{\alpha b,\beta b'}\delta_{b\mu}\delta_{b'\nu},\quad
\eal
where we have defined
\bal
\Lambda_{\alpha b,\beta b'}(k_y) = \sum\limits_l e^{\text{i}k_y l R_y}G_{\alpha\beta}(l R_y \hat y + \br_b - \br_{b'}),\quad
\eal
where $\br_b=x_b\hat x+y_b\hat y$ and $\br_{b'}=x_{b'}\hat x+y_{b'}\hat y$ denote emitter coordinates within a single cell.

Similar to the 2D band structure calculation (see Sec.~1), it is convenient to transform the real-space summation that runs over $l$ to a summation in momentum space. In order to perform this transformation over the lattice sites along the $y$-axis, we need to find an expression for the Green's function that depends on $p_y$ and $x$. Thus, we need to Fourier transform the Green's function along the $y$-axis only. We express the Green's function as
\bal
G_{\alpha\beta}(x,y) = \int \fr{dp_y}{2\pi} g_{\alpha\beta}(p_y;x)e^{\text{i}p_yy},
\eal 
where 
\bal\label{hybridGreens}
g_{\alpha\beta}(p_y;x) = \int \fr{dp_x}{2\pi}g_{\alpha\beta}(p_x,p_y)e^{\text{i}p_xx}.
\eal
In addition, we also make use of Poisson's summation formula in the following form
\bal
\sum\limits_l e^{\text{i}p_y R_y l} = \fr{1}{R_y}\sum\limits_{G_y}2\pi \delta(p_y-G_y),
\eal
where $\{G_y\}$ is the set of reciprocal lattice vectors along the $k_y$ axis. Using these expressions, we obtain
\bal\label{LambdaExp2}
\Lambda_{\alpha b,\beta b'} &=& \fr{1}{R_y}\sum\limits_{G_y}g_{\alpha\beta}(G_y - k_y; x_b-x_{b'})e^{\text{i}(G_y-k_y)(y_b-y_{b'})}\nonumber\\
&=&\fr{1}{R_y}g_{\alpha\beta}(k_y; x_b-x_{b'})e^{-\text{i}k_y(y_b-y_{b'})},
\eal 
where the second equality follows from the fact that only the irreducible Brillouin zone, where $G_y=0$, contributes to the summation and from the observation that $g(-k_y;x)=g(k_y;x)$ due to the time-reversal symmetry of the photonic crystal spectrum.

To proceed, we need to numerically evaluate \refeq{hybridGreens}. Substituting \refeq{GreensMomentum} into \refeq{hybridGreens}, we obtain

\bal
g_{\alpha\beta}(p_y;x) =\mathcal{A}c^2\int\fr{dp_x}{2\pi} \sum\limits_n\fr{E_{\bp,\alpha}^{(n)*}(\br_A)\,E_{\bp,\beta}^{(n)}(\br_A)}{\omega_A^2-\big(\omega_\bp^{(n)}\big)^2}e^{\text{i}p_xx},\nonumber
\eal
where $\bp = (p_x,p_y)$ and the summation runs over all bands. The sum is dominated by the two bands that constitute the Dirac cone (which we label $(+)$ and $(-)$), and there are two inequivalent Dirac cones at the $\mbf K$ and $\mbf K'$ points. Therefore, to a good approximation, we can rewrite the above expression as 
\bal\label{contributions}
g_{\alpha\beta}(p_y;x) &=& g^{(\mbf K,+)}_{\alpha\beta}(p_y;x)+g_{\alpha\beta}^{(\mbf K,-)}(p_y;x)\nonumber\\
&+&  g^{(\mbf K',+)}_{\alpha\beta}(p_y;x)+g_{\alpha\beta}^{(\mbf K',-)}(p_y;x),
\eal
where we have defined
\bal\label{numericalIntegral}
&&g^{(\mbf K,\pm)}_{\alpha\beta}(p_y;x) =\! \mathcal A c^2\!\int \fr{dp_x}{2\pi}\fr{E_{\bp,\alpha}^{(\mbf K,\pm)*}(\br_A)\,E_{\bp,\beta}^{(\mbf K,\pm)}(\br_A)}{\omega_A^2-\big(\omega_\bp^{(\mbf K,\pm)}\big)^2}e^{\text{i}p_xx},\nonumber
\eal
and
\bal\label{numericalIntegral}
&&g^{(\mbf K',\pm)}_{\alpha\beta}(p_y;x) =\! \mathcal A c^2\!\int \fr{dp_x}{2\pi}\fr{E_{\bp,\alpha}^{(\mbf K',\pm)*}(\br_A)\,E_{\bp,\beta}^{(\mbf K',\pm)}(\br_A)}{\omega_A^2-\big(\omega_\bp^{(\mbf K',\pm)}\big)^2}e^{\text{i}p_xx}.\nonumber
\eal
To proceed, we need to utilize the analytic approximations developed in Ref.~\cite{SM_Perczeletal2018} for the band dispersion and electric field near the $\mbf K$ and $\mbf K'$ points. The band dispersion near the $\mbf K$ point is well approximated by 
\bal
\omega_{\bp}^{(\mbf K,\pm)}=\omega_\text{Dirac}\pm v_\text{s}\sqrt{(p_x-p_{\mbf K,x})^2+(p_y-p_{K,y})^2},\nonumber
\eal    
where ${\bp_\mathbf{K}=(p_{\mbf K,x},p_{\mbf K,y})}$
is the quasi-momentum associated with the $\mbf K$ point inside the irreducible Brillouin zone. Similarly, near the $\mbf K'$ point we have
\bal
\omega_{\bp}^{(\mbf K',\pm)}=\omega_\text{Dirac}\pm v_\text{s}\sqrt{(p_x-p_{\mbf K',x})^2+(p_y-p_{\mbf K',y})^2},\nonumber
\eal    
where ${\bp_{\mathbf{K}'}=(p_{\mbf K',x},p_{\mbf K',y})}$.
The electric field of the guided modes near the $\mbf K$ point is well-approximated by
\bal\label{PolarizationStructureK}
\mbf E^{(\mbf K,\pm)}_{\bp}(\mbf r_A) = E_0\left[ \sin\left(\fr{\Phi_{\mbf K}}{2}\mp \fr{\pi}{4}\right)\hat x \pm \sin\left(\fr{\Phi_{\mbf K}}{2}\pm\fr{\pi}{4}\right)\hat y \right]\!,\nonumber
\eal  
where
\bal
\Phi_{\mbf K} (p_x,p_y)= \arctan\left(\fr{p_y-p_{\mbf K,y}}{p_x-p_{\mbf K,x}}\right),\nonumber
\eal
whereas the electric field of the modes near the $\mbf K'$ point is given by
\bal\label{PolarizationStructureKprime}
\mbf E^{(\mbf K',\pm)}_{\bp}(\mbf r_A) = E_0\left[ \sin\left(\fr{\Phi_{\mbf K'}}{2}\pm \fr{\pi}{4}\right)\hat x \mp \sin\left(\fr{\Phi_{\mbf K'}}{2}\mp\fr{\pi}{4}\right)\hat y \right]\!,\nonumber
\eal  
where
\bal
\Phi_{\mbf K'} (p_x,p_y)= \arctan\left(\fr{p_y-p_{\mbf K',y}}{p_x-p_{\mbf K',x}}\right).\nonumber
\eal
The numerical evaluation of $g^{(\mbf K,\pm)}_{\alpha\beta}$ and $g^{(\mbf K',\pm)}_{\alpha\beta}$ is somewhat subtle, as the integrands contain poles, branch cuts and branch points.  Thus, special care has to be taken to define the appropriate integration contour. Here we describe how to evaluate $g^{(\mbf K,\pm)}_{\alpha\beta}$ (i.e. the contributions from the $\mbf K$ point). Evaluating $g^{(\mbf K',\pm)}_{\alpha\beta}$ proceeds analogously. 

\begin{figure}[h!]
\centering
\includegraphics[width=0.45 \textwidth]{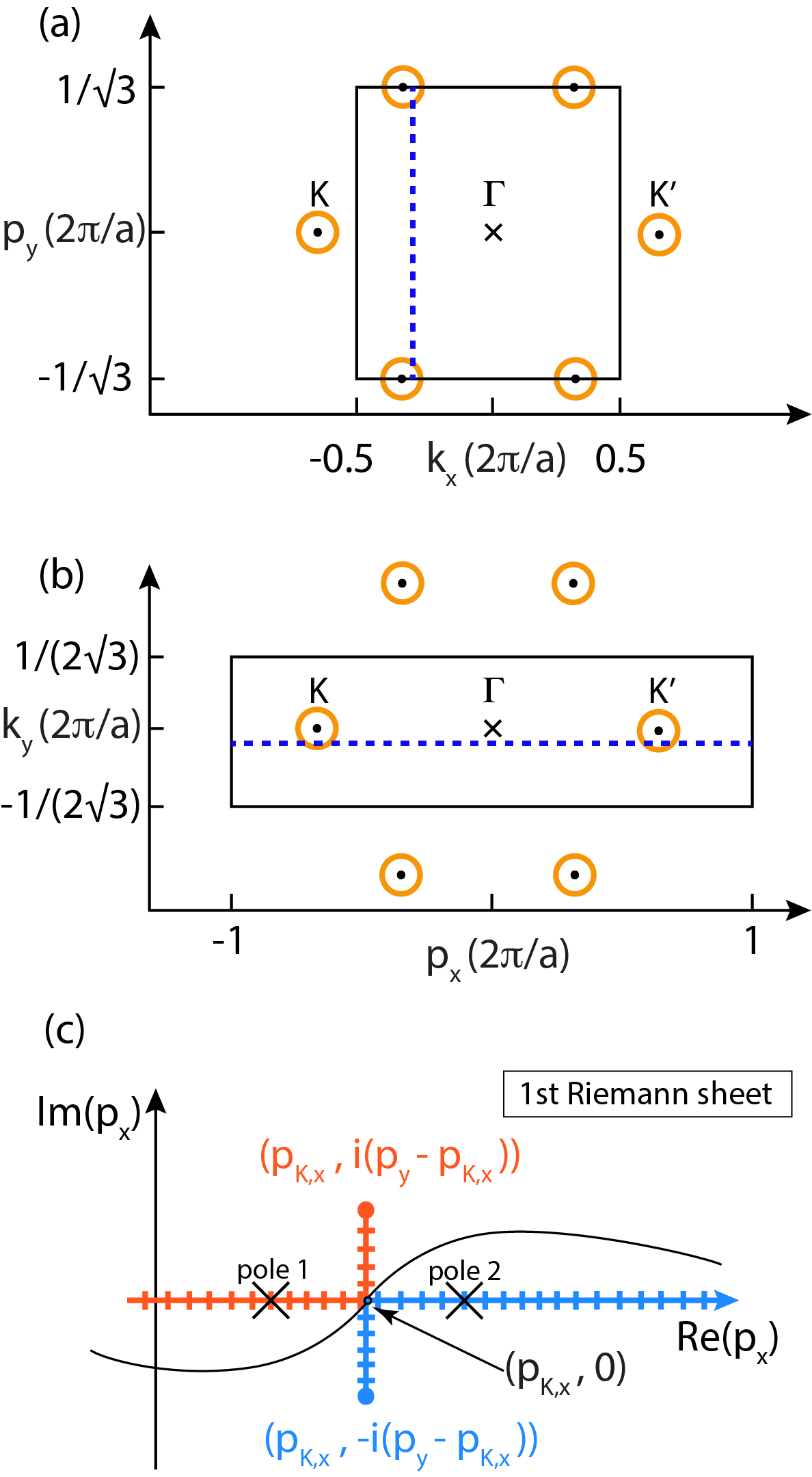}
\caption{
\label{fig:limits}
(a) Integration inside the 2D Brillouin zone to obtain edge states along the $k_x$ axis. Black dots mark the $\mbf K$ and $\mbf K'$ points, whereas orange circles indicate the set of modes in the photonic Dirac cone that are resonant with the emitters. The eigenmodes can be obtained for any Bloch vector $k_x\in[-\pi/a,\pi/a]$, by integrating the interval between $-2\pi/(\sqrt{3}a)\leq p_y \leq 2\pi/(\sqrt{3}a)$. Blue dotted line shows an example of such an integration path. 
(b) Integration inside the 2D Brillouin zone to obtain edge states along the $k_y$ axis. The eigenmodes can be obtained for any Bloch vector $k_y\in[-\pi/(\sqrt{3}a),\pi/(\sqrt{3}a)]$, by integrating the interval between $-2\pi/a\leq p_y \leq 2\pi/a$.  (c) Integration contour near the $\mbf K$ point for the blue dotted line shown in (b). Poles are indicated by black crosses, the branch cuts and branch points the arise from the square root function are shown in red and blue.   } 
\end{figure}

First, note that the integral $g^{(\mbf K,-)}_{\alpha\beta}(p_y;x)$ has a pole when ${\omega_A=\omega_\bp^{(\mbf K,-)}}$, i.e. when the atomic frequency is resonant with the photonic modes of the lower half of the Dirac cone. While these resonant states occupy a circle in the $p_x-p_y$ space, since we are integrating along the $p_x$ axis, there are at most two such singular points along the integration path. These two points are located at
\bal
p_x^\pm = p_{\mbf K, x}\pm \sqrt{\fr{1}{v_\text{s}^2}(\omega_\text{Dirac}-\omega_A)^2-(p_{\mbf K,y}-p_y)^2},\nonumber
\eal
whenever the expression under the square root sign is positive (otherwise there are no poles). These poles in the integrand can be avoided by using an {`$\epsilon$-prescription'}. In particular, we introduce a small imaginary term $+\text{i}\epsilon$ with $\epsilon >0$ into the denominator of the integrand to obtain
\bal
&&g^{(\mbf K,-)}_{\alpha\beta}(p_y;x) = \mathcal A c^2\int \fr{dp_x}{2\pi}\fr{E_{\bp,\alpha}^{(\mbf K,\pm)*}(\br_A)\,E_{\bp,\beta}^{(\mbf K,\pm)}(\br_A)}{\omega_A^2-\big(\omega_\bp^{(\mbf K,\pm)}\big)^2+\text{i}\epsilon}e^{\text{i}p_xx}.\nonumber
\eal
This prescription makes the integral well-defined and corresponds to the {\it causal} Green's function, which represents the outgoing radiation from a point source \cite{SM_Chew1995}. The relevant integration path (that is equivalent to the $\epsilon$-prescription via a contour deformation) is indicated in \reffig{fig:limits}(c). 
Note that for $g^{(\mbf K,+)}_{\alpha\beta}(p_y;x)$ the integrand has no poles. However, we may use the same $\epsilon$-prescription prescription for simplicity.

Furthermore, it is also necessary to choose the appropriate branch when taking the square root $\sqrt{(p_x-p_{\mbf K,x})^2+(p_y-p_{\mbf K,y})^2}$ to find $\omega_{\bp}^{(\mbf K,\pm)}$. The square root function (with a quadratic function as its argument) introduces {\it two} branch cuts that emanate from two branch points at $\pm \text{i}(p_y-p_{\mbf K, y})$, as shown in see \reffig{fig:limits}(c). 
Causality of the Green's function requires that we follow the $\epsilon$-prescription path on the first Riemann sheet \cite{SM_Chew1995}. Doing so corresponds to enforcing 
\bal
\text{Re}(\sqrt{(p_x-p_{\mbf K,x})^2+(p_y-p_{\mbf K,y})^2})\geq 0 \nonumber
\eal
and
\bal
\text{Im}(\sqrt{(p_x-p_{\mbf K,x})^2+(p_y-p_{\mbf K,y})^2})\geq 0. \nonumber
\eal
Note that the two branch cuts come infinitesimally close at $(\text{Re}(p_x),\text{Im}(p_x))=(p_{\mbf K, x},0)$, which makes this point ill-defined. Therefore, it is essential to break the integration contour into separate paths, such that the first path ends at $(p_{\mbf K, x},0)$, whereas the second one starts from there. 

In addition, we also need to ensure that the appropriate branches are chosen when evaluating the inverse tangent function to obtain $\Phi_{\mbf K}$. In particular, we need to ensure that $\Phi_{\mbf K} \in [-\pi,\pi]$.

With these prescriptions, the integral $g^{(\mbf K,\pm)}_{\alpha\beta}(p_y;x)$ can be numerically evaluated (see \reffig{fig:limits}(b) for the relevant integration limits). Analogous prescriptions apply to $g^{(\mbf K',\pm)}_{\alpha\beta}(p_y;x)$. Then, substituting \refeq{contributions} into \refeq{LambdaExp2}, we can diagonalize the matrix $M_{\alpha\beta}(k_y)$ in \refeq{EdgeMatrix} for any ${k_y\in [-2\pi/(2\sqrt{3}a),2\pi/(2\sqrt{3}a)]}$ inside the 1D irreducible Brillouin zone, yielding $2m$ eigenvalues for each $k_y$.

Finally, we note that the calculation for the other stripe orientation can be performed analogously, where the only difference is that the integration is performed along the $p_y$ axis with different limits and the $\mbf K$ and $\mbf K'$ points are located at the edges of the integration interval (see \reffig{fig:limits}(a)). 

\section{Analysis of edge states in the presence of free-space decay} \label{decay}

In the Main Text, we analyzed the topological edges states when emission to free space modes is neglected. Here, we discuss the effect of free-space emission on the decay rate of edge states. 

Before proceeding, we note that standard high-performance calculations for the eigenmodes of photonic crystals, which are based on the plane-wave expansion method to numerically solve Maxwell's equations \cite{SM_Johnson2001}, do not accurately account for the coupling of non-guided modes to free-space photons within the light cone. While there exist techniques for numerically obtaining the out-of-plane decay rate of such modes based on finite-difference time-domain calculations \cite{SM_Fan2002}, doing such calculations at a massive scale would require prohibitively large amounts of computational resources. Therefore, here we utilize prior results from Refs.~\cite{SM_Perczel2017a,SM_Perczel2017b}
on the decay rates of edge states, arising from cooperative atomic behavior, to analyze the edge states of our system in the presence of free-space decay. In particular, we will make use of the observation that edge modes inside the free-space light cone can decay out of plane at a decay rate comparable to the individual free-space linewidth of the atoms $\Gamma_0$, whereas modes outside the light cone cannot couple to free-space photons due to momentum mismatch \cite{SM_Perczel2017a,SM_Perczel2017b}.  

\begin{figure}[h!]
\centering
\includegraphics[width=0.45 \textwidth]{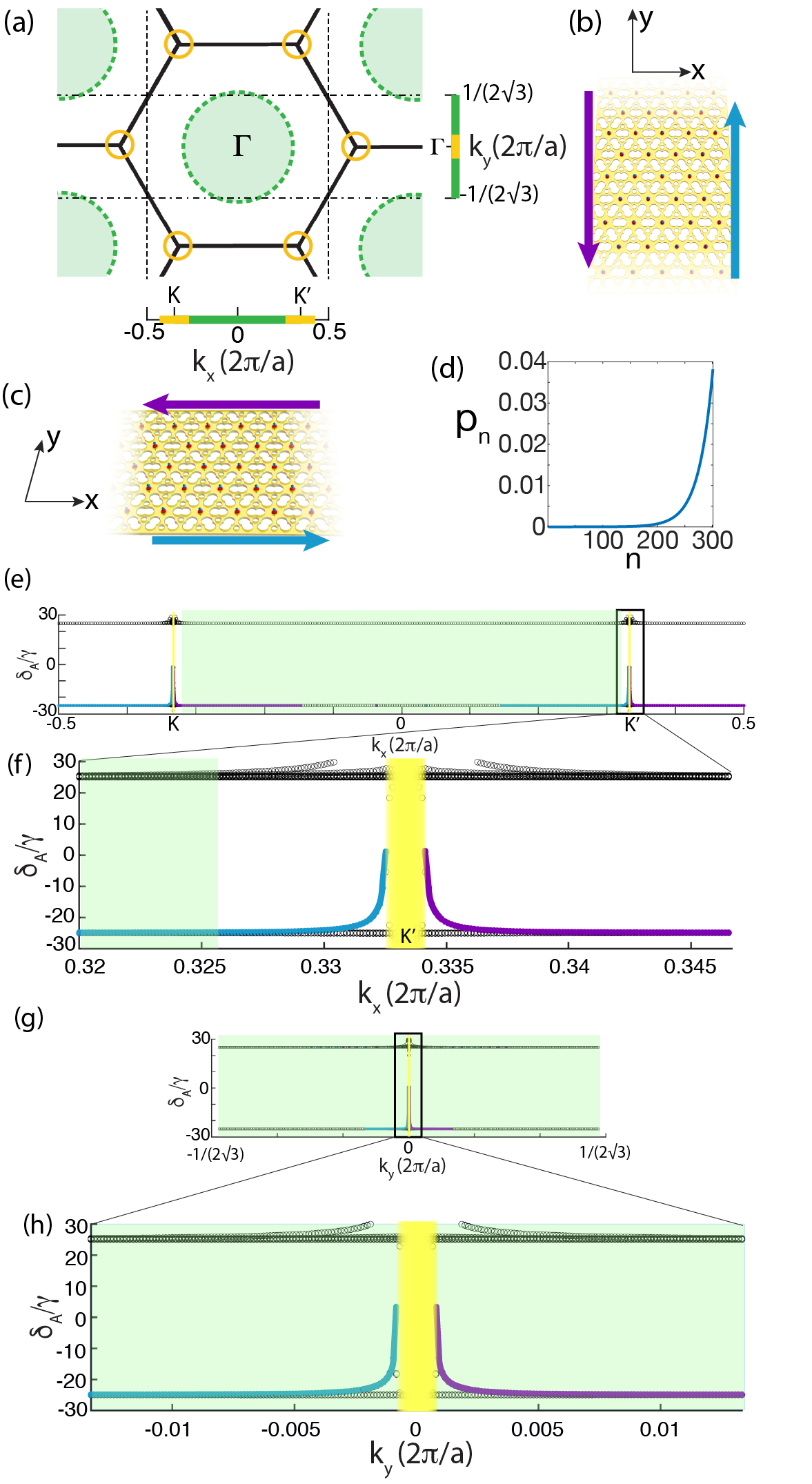}
\caption{
\label{fig:edge_states_SM}
(a) Projection of the 2D Brillouin zone onto the $k_x$ and $k_y$ axes. The light cone region ($|\bk|<2\pi/\lambda$) is shaded green, while guided slab modes resonant with $\omega_A$ are marked with yellow circles. The projection of the light cone spans the entire Brillouin zone on the $k_y$-axis, while covering only the central portion on the $k_x$ axis. (b) \& (c) Schematic illustration of infinite stripes of atoms along the $y$ and $x$ axes, respectively. (d) Typical edge state probability amplitude distribution for a stripe with 300 atoms in the transverse direction, showing localization at one edge. 
(e) Edge states inside the 1D Brillouin zone on the $k_x$ axis. Edge states traverse the gap very close to the $\mbf K$ and $\mbf K'$ points. The lossy region falling inside the light cone is shaded green. (f) Zoomed-in view of the edge states near the $\mbf K'$ point. Edge states on the top (bottom) boundaries are colored purple (blue). Yellow shading indicates region where bands rapidly decay into guided slab modes (the broadening of the modes exceeds the gap size). (d)-(f) Results for a strip along the $y$ axis. The entire Brillouin zone falls insider the light cone making all edge states lossy. } 
\end{figure}

The edge states for the two different stripe orientations are illustrated in detail in Fig.~\ref{fig:edge_states_SM}, where a small detuning of $\delta_A=0.321$THz was used to analyze the edge states for a large gap of $\Delta = 50\gamma$. The properties of the edge states for the two different stripe orientations can be understood by considering the 2D Brillouin zone in momentum space and its projection onto the $k_x$ and $k_y$ axes as shown in Fig.~\ref{fig:edge_states_SM}(a), (b) and (c) (note that we would obtain one of these two types of edge terminations when projecting along any of the 5 other edges or vertices of the 2D Brillouin zone). The light cone region with $|\bk|<2\pi/\lambda$ is shown in green and the modes of the photonic Dirac cone that are resonant with $\omega_A$ are marked with yellow circles. Edge states arise only near the Dirac cones (which are the sources of topology in our system). While the projection of the light cone covers the entire 1D Brillouin zone on the $k_y$ axis, it only covers the central part on the $k_x$ axis. Furthermore, while the two inequivalent Dirac cones project to the center of the Brillouin zone on the $k_y$ axis, on the $k_x$ axis they project outside the light cone. 

Fig.~\ref{fig:edge_states_SM}(d) shows the localization of a typical edge states near the boundary. Fig.~\ref{fig:edge_states_SM}(e) shows the edge states on the $k_x$-axis. Edge states on the top boundary of the strip are colored purple, whereas edge states on the bottom boundary are colored blue. The part of the Brillouin zone that falls inside the light cone ($k_x<2\pi/\lambda$) is shaded green. Modes that fall outside the light cone ($k_x>2\pi/\lambda$) cannot couple to free-space modes due to the momentum mismatch (note that $\left<k_x|k_x'\right>=\delta_{k_x,k_x'}$ in the momentum basis), making these modes long-lived \cite{SM_Perczel2017a,SM_Perczel2017b}. In contrast, modes inside the light cone can couple to free-space photons and, therefore, their decay rate is on the order of $ \Gamma_0$ \cite{SM_Perczel2017a,SM_Perczel2017b}. The edge states have non-negligible dispersion only in the immediate vicinity of the $\mbf K$ and $\mbf K'$ points. Fig.~\ref{fig:edge_states_SM}(f) provides a zoomed-in view of the edge modes near the $\mbf K'$ point. 
Inside the yellow region, atomic bands overlap with guided modes of the Dirac cone with the same energy and momentum, and coupling to these modes effectively dissolves the atomic bands (their broadening exceeds the gap size). Fig.~\ref{fig:edge_states_SM}(f) also shows that the edge states traverse the lower half of the gap outside the light cone and only cross into the yellow region near the center of the gap. Therefore, the edge states in the lower half of the gap are long-lived \cite{SM_Perczel2017a}.

Figs.~\ref{fig:edge_states_SM}(g) and (h) show the edge states on the $k_y$ axis. The edge states have non-negligible dispersion only near the $\mbf K$ and $\mbf K'$ points, which are both projected to the center of the Brillouin zone for this geometry. Thus all edge modes fall inside the light cone and have an out-of-plane decay rate of approximately $\Gamma_0$.

Crucially, for these parameters the edge states traverse an energy interval of $\delta\omega=25\gamma$ over a momentum interval of $\delta k \approx 0.002 (2\pi/a)$, leading to a large group velocity of $v_g\approx \delta\omega/\delta k\approx 2000 a/\gamma^{-1}$. Therefore, making the conservative assumption that $\Gamma_0\approx \gamma$, an excitation would hop 2000 sites before decaying into far-field photons even if carried by edge modes that can decay to free-space photons. 

In general, as $\delta_A$ is decreased to increase the gap size $\Delta$, the group velocity of the edge states increases as $\sim 1/\delta_A^2$. To understand this scaling, note that $v_g\approx \delta \omega/\delta k$, where $\delta\omega$ and $\delta k$ are the energy and momentum intervals traversed by the edge states inside the Brillouin zone. Since $\delta\omega \sim \Delta \sim 1/\delta_A$ and $\delta k\sim \delta_A/v_\text{s}$ (from $\omega_\bp = \omega_\text{Dirac}\pm v_\text{s}|\bk|$), we obtain $v_g\sim 1/\delta_A^2$.

These considerations show that in order to obtain long-lived topological edge states in our system, the lattice edges should be terminated as shown in \reffig{fig:edge_states_SM}(c). For such an edge termination, long-lived edge states can be excited, where losses arise only from finite-size effects, such as corners, and imperfections in the periodic lattice \cite{SM_Perczel2017a}. The key strategy for ensuring that excitations propagate a significant distance, even in the presence of corners and defects, is to maximize the group velocity of the edge states by increasing the gap size. Thus, lossy regions, where decay occurs on a $t\sim \Gamma_0^{-1}$ time scale, will be traversed quickly by the excitation, leading to negligible emission \cite{SM_Perczel2017a}. 

Recall, however, that obtaining a large energy gap requires $\delta_A$ to be small, which leads to a small density of states. Thus, a large system is needed to ensure that there are a sufficient number of edge modes in the gap to enable the dispersive propagation of the excitations on the boundaries. Given the limits on system sizes that can be readily simulated with state-of-the-art computational resources, in the Main Text (Figs. 4 and 5) we focused on systems with a large detuning and small band gap, and ignored emission into free-space modes. However, as the preceding analysis shows (see also Ref.~\cite{SM_Perczel2017a}), the inclusion of emission into free space does not significantly change the results as long as $\Delta\gg \Gamma_0$.


\begin{thebibliography}{78}
\bibitem [{foo()}]{footnote_PC_1}%
  \BibitemOpen
  \href@noop {} {}\bibinfo {howpublished} {Note that all edge state
  calculations are performed for a small gap size (i.e. large $\delta_A$) to
  ensure sufficient density of states even for limited lattice
  sizes.}\BibitemShut {Stop}%
\end{thebibliography}
\end{document}